\documentclass[%
 reprint,
superscriptaddress,
nofootinbib,
eqsecnum,
longbibliography,
 amsmath,amssymb,
 aps,
 prd,
]{revtex4-2}

\usepackage{graphicx}% Include figure files
\usepackage{dcolumn}% Align table columns on decimal point
\usepackage{amsmath}
\usepackage{amssymb}
\usepackage{amsfonts}
\usepackage{mathtools}
\usepackage{url}
\usepackage[dvipsnames]{xcolor}
\usepackage[colorlinks]{hyperref}
\usepackage{float}
\usepackage[inline]{enumitem}
\usepackage{appendix}

\definecolor{burgundy}{rgb}{0.6, 0.0, 0.0}
\definecolor{persianblue}{rgb}{0.11, 0.22, 0.73}

\hypersetup{
    linkcolor=blue,
    citecolor=burgundy,
    filecolor=blue, 
    urlcolor=blue
}
\def\prd{Phys.~Rev.~D}
\def\mnras{MNRAS}
\def\apj{ApJ}

\def\apjs{ApJS}
\def\aap{A\&A}

\def\jcap{JCAP}

\begin{document}

%\title{Changes in cosmic-ray transport properties connect the high-energy features in the electron and proton data}
\title{A diffusive origin for the cosmic-ray spectral hardening reveals signatures of a nearby source in the leptons and protons data}

\author{Ottavio Fornieri}
\affiliation{Deutsches Elektronen-Synchrotron (DESY), Platanenallee 6, D-15738 Zeuthen, Germany}
\affiliation{Department of Physical Sciences, Earth and Environment, University of Siena, Strada Laterina 8, 53100 Siena, Italy}
\affiliation{Instituto de F\'{i}sica Te\'{o}rica UAM-CSIC, Campus de Cantoblanco, E-28049 Madrid, Spain}

\author{Daniele Gaggero}
\affiliation{Instituto de F\'{i}sica Te\'{o}rica UAM-CSIC, Campus de Cantoblanco, E-28049 Madrid, Spain}

\author{Daniel Guberman}
\affiliation{Department of Physical Sciences, Earth and Environment, University of Siena, Strada Laterina 8, 53100 Siena, Italy}
\affiliation{INFN Sezione di Pisa, Polo Fibonacci, Largo B. Pontecorvo 3, 56127 Pisa, Italy}

\author{Loann Brahimi}
\affiliation{Laboratoire Universe et Particules de Montpellier (LUPM) Un. Montpellier, CNRS IN2P3, CC72, place E. Bataillon, 34095, Montpellier Cedex 5, France}

\author{Pedro De La Torre Luque}
\affiliation{The Oskar Klein Centre, Department of Physics, Stockholm University, AlbaNova SE-10691 Stockholm, Sweden}

\author{Alexandre Marcowith}
\affiliation{Laboratoire Universe et Particules de Montpellier (LUPM) Un. Montpellier, CNRS IN2P3, CC72, place E. Bataillon, 34095, Montpellier Cedex 5, France}

%\date{\today}

\begin{abstract}
In this work we aim at reproducing, simultaneously, the spectral feature at $\sim 10 \, \mathrm{TeV}$ in the cosmic-ray proton spectrum, recently reported by the DAMPE Collaboration, together with the spectral break at $\sim 1 \, \mathrm{TeV}$ measured by  H.E.S.S. in the lepton spectrum. Those features are interpreted as signatures of one nearby hidden cosmic-ray accelerator. We show that this interpretation is consistent with the dipole-anisotropy data as long as the rigidity scaling of the diffusion coefficient features a hardening at $\sim 200 \, \mathrm{GV}$, as suggested by the light-nuclei data measured with high accuracy by the AMS-02 Collaboration. Such rigidity-dependent diffusion coefficient is applied consistently to the large-scale diffuse cosmic-ray sea as well as to the particles injected by the nearby source.
\end{abstract}


\maketitle
%\tableofcontents

\section{Introduction}\label{sec:intro}
The past years have witnessed a remarkable increase in the accuracy of both hadronic and leptonic cosmic-ray (CR) data. This advance allowed the community to pinpoint spectral features in primary and secondary species at different energies, which offer a unique opportunity to shed light on the long-standing questions regarding the origin and transport of the non-thermal population of high-energy cosmic particles in our Galaxy~\citep{Gabici:2019jvz, 2018AdSpR..62.2731A}. 
These new data were in particular reported by the AMS-02 Collaboration, which presented some remarkable results. The first general trend is that for primary nuclei (protons, He, C, O) the spectral index progressively hardens at rigidities above $\sim 200 \, \mathrm{GV}$~\citep{PhysRevLett.114.171103}. Recent observations of other primary elements, such as Ne, Mg and Si~\citep{PhysRevLett.124.211102}, confirm this hardening but with smaller breaks. Moreover, the spectral hardening is also found for secondary nuclei (Li, Be, B) but with a value twice as large as the one observed for primary species \citep{PhysRevLett.117.231102, Aguilar:2018njt, 2018PhRvL.121e1103A}. The spectral hardening for the proton spectrum has been confirmed by the DAMPE experiment, which also reported on a spectral {\it softening} at $13.6 \, \mathrm{TeV}$, with the energy index changing from $2.60$ to $2.85$~\citep{An:2019wcw}. This spectral "bump" seems to be firmly established in the observations, being independently measured by the ATIC~\citep{Panov:2011ak} and NUCLEON~\citep{Atkin:2018wsp} experiments.

On the other hand, in the lepton domain, the spectrum has a power-law shape at up to $\sim \mathrm{TeV}$ energies \citep{2014PhRvL.113l1102A} followed by a spectral break at $\sim 1 \, \mathrm{TeV}$, as reported by the H.E.S.S.~\citep{Aharonian:2009ah,kerszberg_ICRC}, CALET~\citep{Adriani:2018ktz} and DAMPE~\citep{Ambrosi:2017wek} collaborations. This spectral feature possibly points towards a nearby old remnant, as shown originally in \citet{Recchia:2018jun} and later elaborated in a wider context in \citet{Fornieri:2019ddi}. Moreover, attempts to assign the high-energy ($E \ge 1 \, \mathrm{TeV}$) observed leptons to known nearby sources --- such as Vela and Cygnus Loop --- using radio data have recently revealed their subdominant contributions (see for example \citet{Manconi:2018azw}).

The physical origins of the different hadronic spectral features, as well as the all-lepton spectra and positron fraction, remain strongly debated. The larger spectral break of the secondary species with respect to primaries suggests a diffusive origin of the effect, as discussed in \textit{e.g.} \citet{2012ApJ...752...68V, Genolini:2017dfb}, whereas the high-energy softening might be due to a contribution of a nearby Supernova Remnant (SNR). However, the latter would require an anomalously slow diffusion in the interstellar medium between the source and the Earth in order to be consistent with the current data on the dipole anisotropy~\citep{Fang:2020cru}. Otherwise, the predicted anisotropy would overshoot the observed data by more than one order of magnitude. %Although this effect is possibly operative, its modelling relies on a good knowledge of the physical properties of the interstellar medium surrounding the source. 
When modelling such effect, a good knowledge of the physical properties of the interstellar medium surrounding the source is required. As a matter of fact, the way CR self-generated turbulence can be damped depends strongly on the type of interstellar phase and in particular its content in neutrals~\citep{2020A&A...633A..72B, 2016MNRAS.461.3552N}. This modelling, including a slow diffusive zone around the source, will deserve a future study. \citet{Liu:2018fjy,Yuan:2020ciu} propose an alternative two-zone disk/halo diffusion setup to reproduce both nuclei and anisotropy spectra. However, this model is not concerned with leptons and has only been applied to background cosmic-ray particles. \citet{2020arXiv201002826M} propose a model interpreting both the spectral hardening and the softening with a CR contribution produced by the reacceleration of the background CR spectrum at a weak, nearby shock, possibly associated with the bow shock of a runaway star. Both spectral components are a consequence of the Earth moving in a magnetic flux tube footed at the shock surface. While drifting in the flux tube, reaccelerated CRs trigger an acoustic instability due to their pressure gradient. The propagation elapsed time is long enough for an Iroshnikov-Kraichan turbulence to be set up and control the CR mean free path. An advantage of this model is that it relies only on two parameters, namely the shock Mach number and the bump rigidity. However, the model does not include any leptonic component, neither it deals with the CR-anisotropy amplitude and phase at $200 \, \mathrm{GV}$ and $13 \, \mathrm{TV}$ properly. \citet{2020FrPhy..1624501Y} also consider a nearby source to explain both the hardening and softening spectral features, though not discussing the leptons. To that aim, the model assumes a spatially dependent diffusion coefficient for the background CR-sea but a single-power-law diffusion coefficient for the source components. Even though this could be motivated by the fact that high-energy particles diffuse much faster than low-energy ones, eventually spending most of their time in one zone only, this effect is quantified in our modelling and it is found to be true only for energies above $E \sim 10 \, \mathrm{TeV}$. This is in fact the source of the problems in the all-lepton spectrum, where a particle population is required to contribute also in the range $E \sim [50 \, \mathrm{GeV}, \, 1 \, \mathrm{TeV}]$. Interestingly, though, the authors show that a different elemental composition at the source may explain the spectral difference between the He, C, O and the Ne, Mg, Si groups.

In this paper (and in the associated Letter \cite{Fornieri:2021PRL}), we propose a comprehensive scenario that correctly reproduces proton, lepton and anisotropy spectra.

Our model features two key points of novelty. 
\begin{enumerate*}[label=(\roman*)]
\item First, we argue that a nearby, possibly hidden, old Supernova Remnant is responsible for both the hadronic {\it bump} measured by DAMPE/NUCLEON/ATIC and the leptonic break reported by H.E.S.S.
\item We then consider, for the first time in the background+source context, a transport scenario featuring a rigidity scaling that progressively hardens --- deviating from the single power-law --- as suggested by AMS-02 light nuclei data. We show that this crucial ingredient allows us to reproduce the anisotropy data.
\end{enumerate*}

The paper layout is as follows. In Section \ref{sec:our_transport_setup}, we describe our transport model, with particular attention to the phenomenological treatment that implements a variable slope of the diffusion coefficient for the nearby-source solution as well. In Section \ref{sec:nearby_source_contributions}, we characterize the contributions from a hidden nearby source, connecting for the first time the leptonic and hadronic features and showing that those interpretations are consistent with the CR dipole anisotropy. Finally, in Section \ref{sec:discussion} and \ref{sec:conclusions}, we discuss the results and derive our conclusions.

\section{Our transport setup}\label{sec:our_transport_setup}

In this section, we describe the propagation setup that will be used throughout the paper, which is based on the model settings presented in \citet{Fornieri:2019ddi}. 

We consider a large-scale diffuse background of hadronic and leptonic cosmic particles, plus a contribution from a nearby accelerator.
While the latter component is computed in a semi-analytical way, the former (\textit{i.e.} a smooth contribution) is characterized by solving the general diffusion-loss transport equation with the {\tt DRAGON}\footnote{\url{https://github.com/cosmicrays/DRAGON}}~\citep{Evoli:2008dv,2017JCAP...02..015E} numerical code. {\tt DRAGON} takes into account all the physical processes from low-energy up to high-energy effects. In this work, we consider a $2$D configuration, with cylindrical symmetry and an azimuthal-only Galactic magnetic field geometry. The physical ingredients implemented in our run, for what concerns the environment setup --- the gas distribution, interstellar radiation field and intensity of the regular and turbulent magnetic fields --- as well as the CR physics parameters --- the CR source distribution and the non-adiabatic energy losses suffered by both hadrons and leptons --- are the same described in \citet{Fornieri:2019ddi}.

However, a key difference with respect to the aforementioned work resides in the assumption on the diffusion coefficient. As mentioned in the introduction, the more pronounced effect detected in the purely secondary species seems to point towards a feature in the transport. Specifically, the CR distribution function at the disk level, that is found by solving the transport equation, can be written as $f_0 (E) \sim S(E) / D(E)$, where $S(E)$ is the particle injection-spectrum and $D(E) \sim E^{\delta}$ the diffusion coefficient. For primary species, $S(E) \sim E^{-\Gamma_{\mathrm{inj}}}$, from which we get $f^{\mathrm{pri}}_0 \sim E^{-\Gamma_{\mathrm{inj}} - \delta}$, while for secondaries, the injection spectrum is the propagated spectrum of the primaries, resulting in $f^{\mathrm{sec}}_0 \sim E^{-\Gamma_{\mathrm{inj}} - \delta} / D(E) = E^{-\Gamma_{\mathrm{inj}} - 2\delta}$. This implies that any change in the slope of the diffusion coefficient will produce a change in the secondaries' spectrum that is twice as large as that in the primaries. This is what is observed by AMS-02~\citep{Aguilar:2018njt} for the CR hardening at $\sim 200 \, \mathrm{GeV}$.

As a consequence, assuming this hardening to be of diffusive origin, it appears quite natural that equal changes in the transport properties should affect the propagation of particles from nearby sources as well.

To consider this, we study the phenomenological setup considered in \citet{Tomassetti:2012ga}, where the slope of the diffusion coefficient --- typically parametrized as $D(E) = D_0 \left( \frac{E}{E_0} \right)^{\delta(E)}$, with $D_0$ normalization at reference energy $E_0$ and $\delta$ here changing with $E$ --- smoothly hardens as energy (or, equivalently, rigidity) increases, assuming the following expression:
\begin{equation}\label{eq:delta_tomassetti}
    \frac{d \log D(\rho)}{d \log \rho} \equiv \gamma(\rho) \approx \gamma_{\mathrm{high}} + \frac{\Delta}{1 + \frac{\xi}{1 - \xi} \left( \frac{\rho}{\rho_0} \right)^{\Delta}},
\end{equation}
where $\rho$ is the particle rigidity, $\rho_0$ is the reference rigidity and ($\gamma_{\mathrm{high}}, \Delta, \xi$) are free parameters of the model.

\begin{figure}[ht]
    \centering
    \includegraphics[scale=0.28]{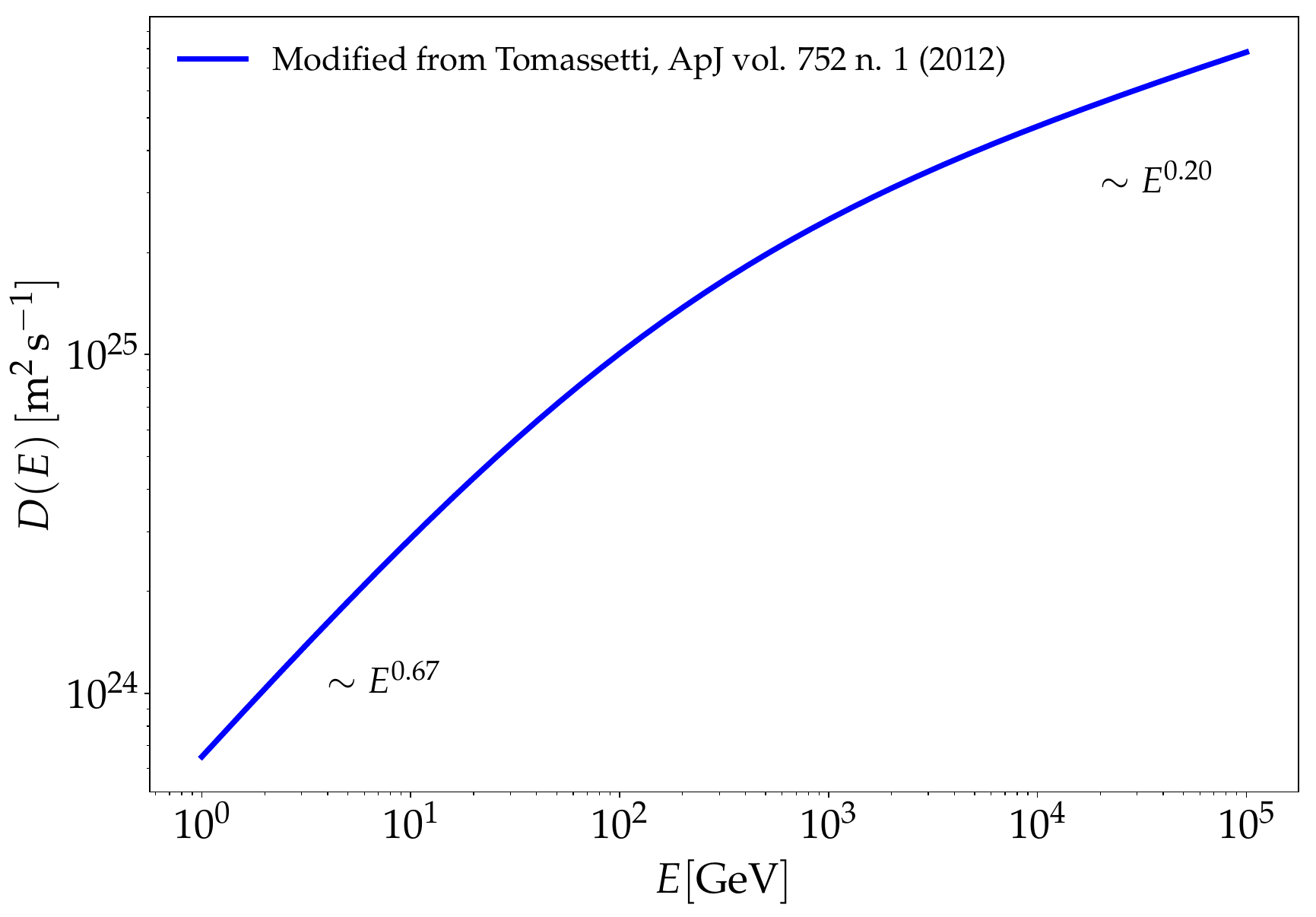}
    \caption{\small{The diffusion coefficient obtained from the parametrization $D(E) = D_0 \left(\frac{E}{E_0} \right)^{\delta(E)}$, modifying the parameters $(\gamma_{\mathrm{High}}, \Delta)$ starting from the {\tt THMb} model in \citet{Tomassetti:2012ga}, as described in the text.}}
    \label{fig:Diffusion_coefficient_Tomassetti}
\end{figure}

In order to account for the nearby-source contribution to the proton flux, we slightly modify the parameters $(\gamma_{\mathrm{High}}, \Delta)$ starting from their {\tt THMb}-model (\textit{Two-Halo Model b}) values. In what follows, they become $\gamma_{\mathrm{high}} = 0.19$, $\Delta = 0.53$, while the others are left unchanged, as $\xi = 0.1$, with a normalized diffusion coefficient $D_0 = 1.21 \cdot 10^{28} \, \mathrm{cm^2 \, s^{-1}}$ at reference rigidity $2 \, \mathrm{GV}$. With these parameters, the diffusion coefficient presents a smooth transition, specifically as shown in Figure \ref{fig:Diffusion_coefficient_Tomassetti}.

The key point shown in \citet{Tomassetti:2012ga} is that such a setup is formally equivalent to a two-zone transport model featuring a change in the properties of the interstellar medium (ISM) between an inner-halo ($|z| < \xi L$) region and an extended-halo ($\xi L < |z| < L$) region, where $L \sim 4 \, \mathrm{kpc}$ and $\xi \sim \mathcal{O}(0.1)$. Possible underlying physical explanations for this change of slope in the diffusion coefficient have been proposed.
\begin{enumerate*}[label=(\roman*)]
    \item\label{item:Blasi_hardening} One assigns it to the transition between a diffusion regime, generated by self-generated turbulence, to another one for which an external large-scale $(L_{\mathrm{inj}} \sim 10-100 \, \mathrm{pc})$ Alfvénic cascade is responsible~\citep{Blasi:2012yr}.
    \item\label{item:Yan&Lazarian_hardening} Alternatively, based on the theory developed in a relatively recent series of papers~\citep{Yan:2002qm, Yan:2004aq, Yan:2007uc}, this change is interpreted within a framework where different damping mechanisms, dominating in the two regions, produce a different behaviour in the turbulent waves, which are the scattering centers that cause CR diffusion~\citep{Evoli:2013lma}.
\end{enumerate*}

It has been shown recently that the real picture could involve a combination of the two hypothesis above~\citep{10.1093/mnras/stab355}: in fact, taking into account all the three MHD modes in which an external turbulent cascade is decomposed, the role of Alfvén modes in shaping the $D(E)$ is significantly subdominant, emerging in turn the effect of the fast modes that generate a diffusion coefficient that is likely not a single power-law of the rigidity. However, under plausible physical condition, has not been possible to reproduce the CR spectra in the whole energy range and good results have been found in the high-energy regime $(E > 200 \, \mathrm{GeV})$, in terms of particle spectra as well as secondary/primary ratios. With this regard, it is worth mentioning that the energy $E \sim 200 \, \mathrm{GeV}$ at which self-generated turbulence stops dominating the CR diffusion was already predicted in \citet{Farmer:2003mz}.

In \citet{Tomassetti:2012ga}, the transport equation is analytically solved under simplifying conditions and the diffusion parameters of Equation \ref{fig:Diffusion_coefficient_Tomassetti} are adjusted to the B/C data available at that time. Later, the same author found a better agreement to the updated observations by incorporating a factor $\beta^{\eta}$ (where $\beta = v/c$ and $\eta \sim -0.4$) into the definition of the diffusion coefficient, in \citet{Feng:2016loc}. This change in the low energy trend of $D(E)$ is discussed in \citet{2019PhRvD..99l3028G}. It has been interpreted in terms of dissipation of magneto-hydrodynamic (MHD) waves in the interstellar plasma~\citep{Ptuskin:2005ax} or, alternatively, considering non-resonant interactions between the cosmic rays and the same turbulent waves~\citep{Reichherzer:2019dmb}. As it is clear, adding this factor has a negligible effect at particle energies for which $\beta \rightarrow 1$, therefore it can be safely ignored in the computation of the spectra from our isolated nearby source.

\begin{figure}[ht]
    \centering
    \includegraphics[scale=0.28]{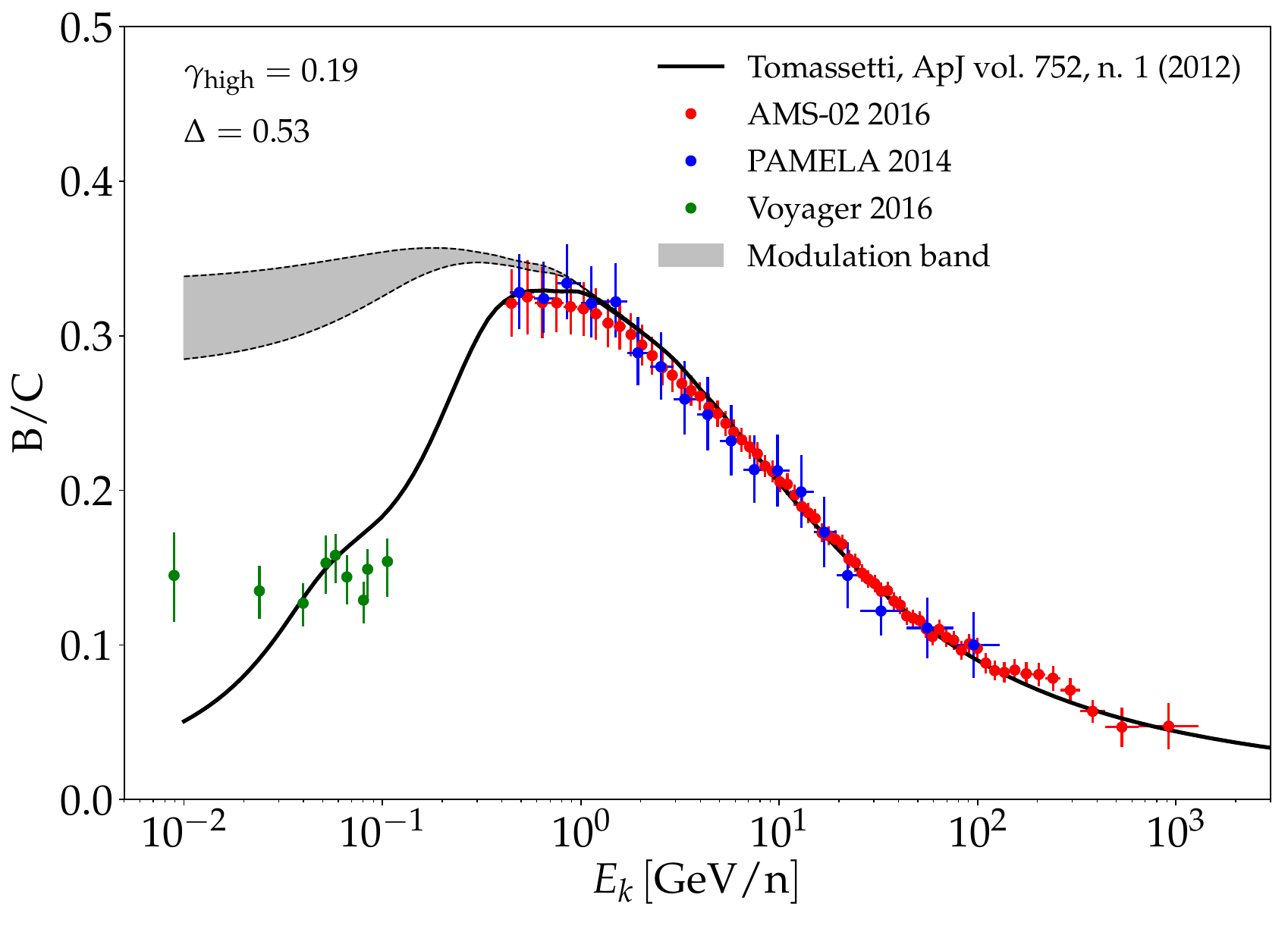}
    \caption{\small{B/C ratio computed for the described model with the {\tt DRAGON} numerical solver with and without adding the influence of the solar modulation, against AMS-02 (red) and PAMELA (blue) data points. Voyager unmodulated data points are also shown at low energy (green). References are in the text.}}
    \label{fig:B_over_C_Tomassetti}
\end{figure}

Here, as mentioned above, we solve the equation for the large-scale background with the {\tt DRAGON} numerical solver, that takes into account all the
processes approximated in the analytical solution. As shown in Figure \ref{fig:B_over_C_Tomassetti}, we find that the B/C flux-ratio observed by AMS-02~\citep{PhysRevLett.117.231102} and PAMELA~\citep{Adriani:2014xoa} can be nicely reproduced using $\eta=-0.5$ and adjusting the values of $\gamma_{\mathrm{high}}$ and $\Delta$ to 0.19 and 0.53, respectively, as indicated in the figure. The Voyager-1~\citep{Cummings_2016} data points, measured outside of the heliopause, are captured at low energy by our unmodulated black solid line. The solar modulation is taken into account using the \textit{force-field} approximation~\citep{1968ApJ...154.1011G}, with an effective potential $\langle \phi_{\mathrm{mod}} \rangle = 0.54 \pm 0.10$~\citep{2005JGRA..11012108U,2011JGRA..116.2104U}. In the plot, its effect is shown as a grey band. We highlight that this framework suitably reproduces the high-energy range of the B/C observations as well as the hardening found in the primary CR-species, for which the model was originally built.

In this paper, the transport setup described in Equation \eqref{eq:delta_tomassetti} --- and shown in Figure \ref{fig:Diffusion_coefficient_Tomassetti} --- is adopted consistently in both the large-scale propagation and in the propagation of particles from the nearby remnant. As it will be shown below, this ingredient plays a key role in reconciling the high-energy break in the all-lepton spectrum ($E_{e^{\pm}} \sim 1 \, \mathrm{TeV}$) with the \textit{bump} recently reported by DAMPE in the proton spectrum at $E_{p} \sim 10 \, \mathrm{TeV}$. Besides, it is crucial to correctly reproduce the cosmic-ray dipole anisotropy data.

\section{Towards a consistent picture of electron, proton and anisotropy data}\label{sec:nearby_source_contributions}

It has been mentioned in the introduction that particles coming from observed nearby sources cannot account for most of the measured high-energy leptons. However, it is natural to wonder whether it is plausible to invoke only one additional hidden source or rather a plurality of them. An answer, with a detailed estimation, is given in Appendix \ref{app:numberRemnants}. In fact, based on the rate of Supernova events in the Galaxy~\citep{Ferriere:2001rg} --- the same implemented in {\tt DRAGON} --- and on the massive losses that leptons undergo during the journey towards the Earth, we find that we expect $N_{\mathrm{SNR}} \sim 2$ Supernova explosions in the vicinity of the Solar system. The catalogues already list more than five~\citep{Ferrand:2012jh}, which however have been found not to contribute to the propagated leptons~\citep{Fornieri:2019ddi, Manconi:2018azw}. Hence, we conclude that considering only one hidden source is a physically well-motivated choice.

Therefore, within the transport setup presented above, we discuss here a scenario based on the contribution from an old, hidden Supernova Remnant as a time-dependent source of cosmic electrons and protons. 

In order to match the proton and lepton fluxes, the accelerator we are considering is characterized by distance $d = 300 \, \mathrm{pc}$ and age $t_{\mathrm{age}} = 2 \cdot 10^5 \, \mathrm{yr}$. The age is fixed by the location of the $\sim 1 \, \mathrm{TeV}$ spectral break in the lepton flux, based on the traveling time of the released particles that are affected by massive losses, as $\Delta t_{\mathrm{travel}} \sim 1 \big/\left( b_0 E_{\mathrm{obs}} \right) \sim 10^5 \, \mathrm{yr}$, with $b_0 \simeq 1.4 \cdot 10^{-16} \, \mathrm{GeV}^{-1} \cdot \mathrm{s}^{-1}$ for standard ISM conditions and regular magnetic field $B_0 = 3 \, \mu\mathrm{G}$. In fact, $t_{\mathrm{age}} = t_{\mathrm{rel}} + \Delta t_{\mathrm{travel}}$, where $t_{\mathrm{rel}}$ is the release time, which is of course $t_{\mathrm{rel}} \leq \Delta t_{\mathrm{travel}}$. Since leptons of such high-energy can only be accelerated and injected in the early stages of the SNR evolution --- according to Equations \eqref{eq:timescales_SNR_STAGES}, $\sim 1 \, \mathrm{TeV}$ electrons are released at $\sim 10^4 \, \mathrm{yr}$ --- then $t_{\mathrm{age}} = t_{\mathrm{rel}} + \Delta t_{\mathrm{travel}} \approx \Delta t_{\mathrm{travel}}$. Such rough estimation shows that the reference age is determined with high level of robustness by the loss mechanism, and that no degeneracy holds with the source distance, that does not enter the calculation. The distance we chose, on the other hand, is the one that better reproduced the data in the region around the break. As studied in \citet{Fornieri:2019ddi}, in fact, the peak of the propagated spectra broadens as the source-to-observer distance decreases. An \textit{a posteriori} check will confirm that the total energy budget of such accelerator and the relative conversion efficiencies into accelerated protons and electrons are compatible with theoretical expectations. We assume that particles remain confined inside the SN shock as long as their energy is lower than the maximum allowed value --- we refer to this value as \textit{escape energy}. This implies an energy-dependent release time that is regulated by the different stages of the SNR evolution and is different for protons and electrons. In this work, we assume that the CR escape energy is dominated by the limited current that particles can generate to trigger non-resonant streaming instability during the \textit{free expansion} and \textit{Sedov} phases; conversely, it is limited by geometrical losses during the later radiative phases. The time scale of each phase of the SNR evolution, as well as the details to compute the escape energy at each time instant, are discussed in Appendix \ref{app:release_time}. With this regard, we remark that the parameters of the ISM that impact the release time are not the result of a fitting procedure, but rather they are chosen based on independent observations.

After the escape, particles are injected into the ISM according to a time-dependent luminosity function $L(t)$ and transported from the source to the Earth via the following diffusion-loss equation, written in polar coordinates~\citep{1995PhRvD..52.3265A}:
\begin{equation}\label{eq:transport_equation_polar_coordinates}
\begin{split}
    \frac{\partial f(E,t,r)}{\partial t} &= \frac{D(E)}{r^2} \frac{\partial}{\partial r} r^2 \frac{\partial f}{\partial r} +\\
    &+ \frac{\partial}{\partial E} (b(E) f) + Q(E,t,r),
\end{split}    
\end{equation}
where $Q(E,t,r) = S(E)L(t)\delta(r)$ is the source term --- with the luminosity function of the form $L(t) = L_0 \Big/ \left( 1 + \frac{t}{\tau_{\mathrm{d}}} \right)^{\alpha_{\mathrm{d}}}$, where $\tau_{\mathrm{d}} = 10^5 \, \mathrm{yr}$ and $\alpha_{\mathrm{d}} = 2$ ---, $D(E) = D_0 \left( \frac{E}{E_0} \right)^{\delta(E)}$ --- as described in the previous section --- and $b(E) \equiv \frac{dE}{dt}$ is the rate of energy-loss, that depends on the specific particles we are considering. 

The previous equation neglects low-energy effects such as advection and reacceleration, since, in the energy regime we are interested (above $\sim 1$ GeV), the transport process is nearly completely diffusive.

Finally, as a consistency check, the total energy budget associated to each CR population injected by the source can be calculated as follows:
\begin{equation}\label{eq:total_energy_budget}
    E_{\mathrm{tot}} = \int dr  \int^{t_{\mathrm{age}}}_{t_{\mathrm{rel}} (E)} dt \int_0^{+ \infty} dE \, E \cdot Q(E, t, r),
\end{equation}
where clearly $t_{\mathrm{rel}}$ is the instant of the release and $t_{\mathrm{age}}$ the current age of the source.

\subsection{All-lepton spectrum}\label{subsec:all_lepton_spectrum}
In the case of leptonic cosmic rays above $\sim 1 \, \mathrm{GeV}$, the energy-loss term accounts for Inverse Compton (IC) scattering and synchrotron losses. The IC cross-section above $\sim 50 \, \mathrm{GeV}$ gets modified by relativistic effects, as shown in \citet{2017PhRvD..96j3013H}, and the loss rate can be written as follows:
\begin{equation}\label{eq:losses_KN}
    b(E) = -\frac{4}{3} \, c \sigma_T \left[ f^{i}_{\mathrm{KN}} U_i + U_B \right] \left( \frac{E}{m_e c^2} \right)^2
\end{equation}
where $\sigma_T \simeq 6.65 \cdot 10^{-25} \, \mathrm{cm}^2$ is the Thomson cross-section, $(U_i, \, U_B)$ are respectively the energy density of the Interstellar Radiation Field (ISRF) components and of the background magnetic field, and  $f^i_{\mathrm{KN}}$ is the approximated correction factor:
\begin{equation}\label{eq:f_KN}
f^{i}_{\rm KN}(E) \simeq \frac{\frac{45}{64 \pi^2} \cdot (m_e c^2/ k_{\rm B}T_i)^2}{ \frac{45}{64 \pi^2} \cdot  (m_e c^2/ k_{\rm B} T_i)^2 + (E^2 / m_e^2 c^4)},
\end{equation}
where $T_i$ are the black-body spectrum temperatures corresponding to $U_i$. For each contribution, we adopted the reference value reported in \citet{2020arXiv200701302E}.

The Green function of Equation \eqref{eq:transport_equation_polar_coordinates} reads:
\begin{equation}\label{eq:solution_transport_equation_implicit}
    f(r,t,E) = \frac{Q(E_{\textrm{t}}) b(E_{\textrm{t}})}{\pi^{3/2} b(E) r^3_{\textrm{diff}}} \cdot e^{-\frac{r^2}{r^2_{\textrm{diff}}}},
\end{equation}
where $E_{\textrm{t}}$ refers to the energy at a time $(t - t_{\textrm{rel}})$ ago and $r^2_{\textrm{diff}} (E_{\textrm{t}},E) \equiv  4 \int_{E_{\textrm{t}}}^{E} \frac{D(E')}{b(E')} dE'$ is the square of the diffusive distance travelled by a particle loosing its energy from $E_{\textrm{t}}$ to $E$. This solution is still general, in that it does not contain any information about the injection term.

The dependence of the diffusion slope on energy has to be included in the integral giving the diffusive distance $\sqrt{r^2_{\mathrm{diff}}}$, as follows:
\begin{equation}\label{eq:diffusive_distance_variable_delta}
\begin{aligned}
    r^2_{\mathrm{diff}} (E_{\mathrm{t}}, E) &= 4 \int_{E_{\mathrm{t}}}^E \frac{D_0 \left( \frac{E'}{E_0} \right)^{\delta(E')}}{b(E')} \, dE' \\
    &= 4 D_0 E_0 \int_{\omega_{\mathrm{t}} = E_{\mathrm{t}}/E_0}^{E/E_0} \frac{\omega^{\delta(E_0\omega)}}{b(E_0 \omega)} \, d \omega,
\end{aligned}
\end{equation}
where the last step is justified by the simple change of variable $\omega = \frac{E'}{E_0}$. In lack of an analytic function $\delta(\omega)$, the integral can be solved numerically.

As a last step, to obtain the propagated spectra at Earth, we have to integrate Equation \eqref{eq:solution_transport_equation_implicit} over time, from the instant of the release from the source to the current time, featuring a model for the time evolution of the luminosity. This is discussed in details in \citet{1995PhRvD..52.3265A}, and summarized in \citet{Fornieri:2019ddi} (their Appendix A).
\begin{figure}[ht]
    \centering
    \includegraphics[scale=0.28]{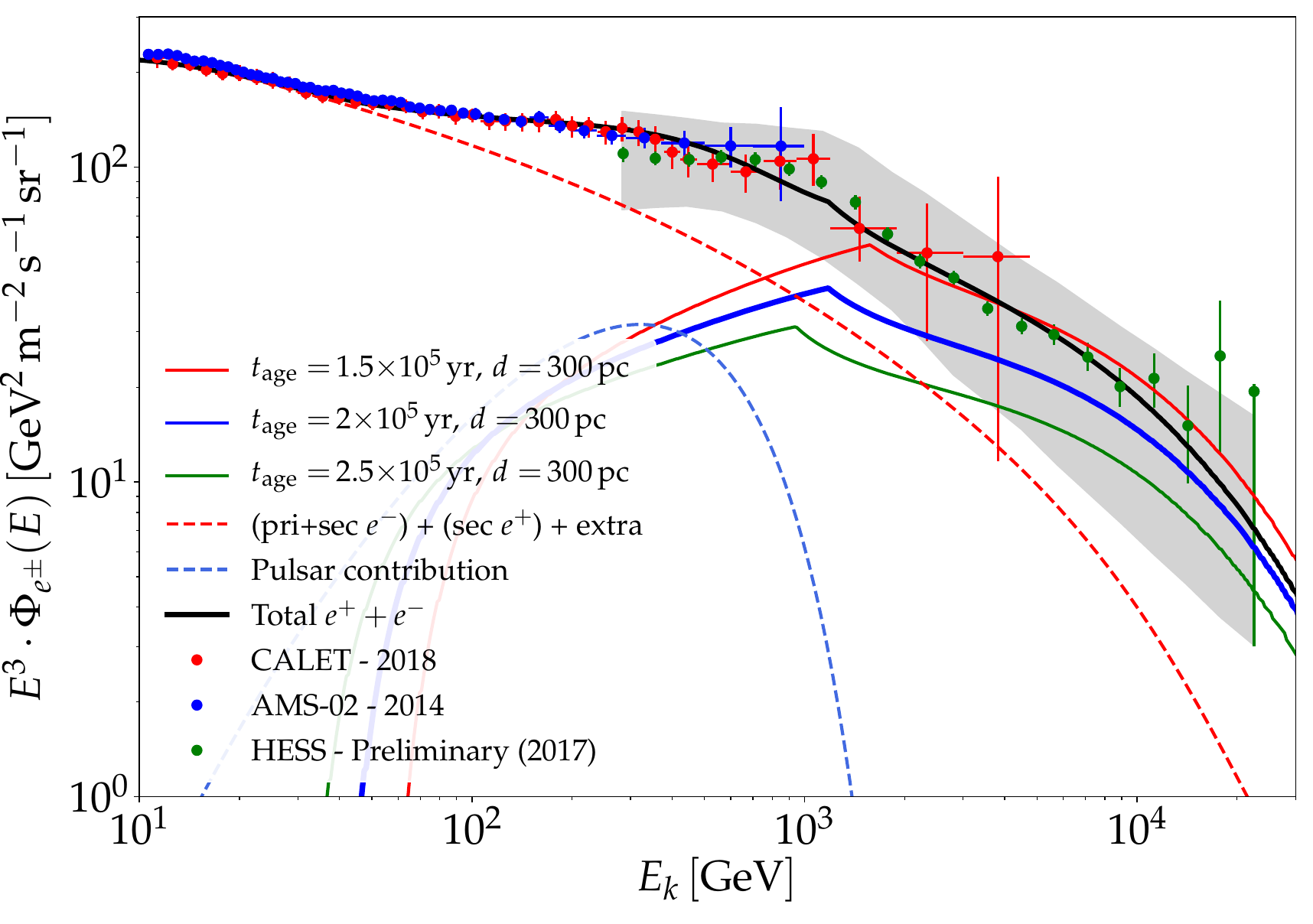}
    \caption{\small{The all-lepton spectrum as the sum of a smooth background of primary $e^-$ + secondary $e^{\pm}$ + extra $e^{\pm}$ (red dashed line), a fit of the positron flux (blue dashed line) and the single-source contribution calculated in this work for the corresponding age $t_{\mathrm{age}} = 2 \cdot 10^5 \, \mathrm{yr}$ (blue solid line). Other ages (red and green solid lines) are added for comparison. The grey band reports the systematics for the H.E.S.S. data points.}}
    \label{fig:all_lepton_solutions_different_ages}
\end{figure}

In Figure \ref{fig:all_lepton_solutions_different_ages} we show the $e^{+} + e^{-}$ propagated spectrum resulting from the convolution of several components, plotted against data from AMS-02~\citep{Aguilar:2014fea}, CALET~\citep{Adriani:2018ktz} and H.E.S.S.~\citep{kerszberg_ICRC}. Data from other experiments have not been added to avoid superposition, being consistent with the present ones. The smooth diffuse background (red dashed line) is the sum of:
\begin{enumerate*}[label=(\roman*)]
    \item\label{item:primary_electrons} primary $e^{-}$, injected with {\tt DRAGON} with a power-law spectrum $\Gamma^{{\tt DRA} \, e^{-}}_{\mathrm{inj}} = 2.74$ and a cutoff $E^{{\tt DRA} \, e^{-}}_{\mathrm{cut}} = 20 \, \mathrm{TeV}$ that is estimated equating the acceleration and loss timescales~\citep{Vink:2011ei};
    \item\label{item:secondary_leptons} secondary $e^{\pm}$, fixed by the {\tt DRAGON}-propagated primary species;
    \item\label{item:extra_component} a smooth {\it extra-component} of primary $e^+ + e^-$ pairs, that represents the convolution of a large ($\sim \mathcal{O}(10^4)$) number of old ($t_{\mathrm{age}} > 10^6 \, \mathrm{yr}$) pulsars (see \citet{Fornieri:2019ddi}).
\end{enumerate*}
For what concerns their relative contribution, we note that such \textit{extra component} is fixed by fitting the low-energy ($E_{e^{+}} < 40 \, \mathrm{GeV}$) positron flux (see \citet{Fornieri:2019ddi}), therefore its impact amounts to just $\sim 3 \%$ of the total leptonic flux, while the secondary component is even smaller, up to $\sim 1 \%$.

The solar modulation is ignored in the plot, as it nearly has no effect at such energies ($E \geq 10 \, \mathrm{GeV}$).

The blue dashed curve represents a fit of the positron flux: here, we invoke pulsars and use the fit to the AMS-02 data points performed in \citet{Fornieri:2019ddi}, for the simplest case of a \textit{burst-like} injection and an intrinsic cutoff in the injection spectrum. Other parametrizations of the positron component do not change the final contribution, as they are fit over the positron flux, expected to originate from a separate class of sources, regardless of the physical nature.

The three solid curves correspond to the contribution from the hidden remnant discussed in this work. They are computed by solving Equation \eqref{eq:transport_equation_polar_coordinates} for different ages, with the calculations described above in this section. The electron population is injected as a single power-law with a slope $\Gamma_{\mathrm{inj}}^{e^-} = 2.45$. This spectrum is softer than the one used for the proton flux, as we will see in the next section. However, such difference is physically motivated by the sychrotron losses that electrons undergo before being released~\citep{Diesing:2019lwm}. The total energy budget associated to the leptonic population, computed by means of Equation \eqref{eq:total_energy_budget}, is $E^{e^-}_{\mathrm{tot}} \simeq 4.5 \cdot 10^{47} \, \mathrm{erg}$.

Finally, the black curve is the sum of all the contributions, where we have chosen the source of age $t_{\mathrm{age}} = 2 \cdot 10^5 \, \mathrm{yr}$ as our best-fit choice (blue solid). 

The plot shows how the energy-dependent release cuts off the low-energy particles ($E \lesssim 100 \, \mathrm{GeV}$) that --- being the last ones to reach the shock escape energy --- did not have the time to be released and then propagate to the Earth. This effect is amplified by the KN correction. Indeed, a corrected cross-section increases the propagated flux of a factor $\sim 1.5-2$, with respect to the non-relativistic treatment, above energies $E \sim 200 \, \mathrm{GeV}$~\citep{2020arXiv200701302E}. Therefore, in order to reproduce the $\sim 1 \, \mathrm{TeV}$ peak, a lower injected flux is needed.

As far as the luminosity function is concerned, its parameters $(\alpha_d, \tau_d)$ are found in order to match the observed lepton flux, although we studied how our predictions change as they vary. In particular, we considered $\alpha_{\mathrm{d}} \in [1, \, 3]$ and reported negligible variations in the spectrum. On the other hand, while varying $\tau_{\mathrm{d}}$ in the range $[10^4, \, 2 \cdot 10^5] \, \mathrm{yr}$ does not qualitatively change the results, smaller values cannot reproduce the data points above the $\sim \, \mathrm{TeV}$ break. Indeed, since $\tau_{\mathrm{d}}$ acts as a timescale for the luminosity function, a quickly decaying luminosity would approach the limit of a burst-like injection ($L(t) \rightarrow L_0 \, \delta(t - t_{\mathrm{rel}}) \, dt$), and accordingly the $\sim \, \mathrm{TeV}$ peak energy allowed by the source age would be followed by an abrupt cutoff in the spectrum. This leads us to conclude that a declining luminosity from the source is necessary to match the observations.

\subsection{Proton spectrum}

The proton data are characterized by a hardening at $\sim 200$ GeV and a softening at energies as high as $\sim 13$ TeV. Here, we connect this feature to the same hidden remnant considered in the previous section.

To compute the contribution from the nearby source to the proton flux, we use again Equation \eqref{eq:transport_equation_polar_coordinates}, neglecting the loss processes considered for leptons, as they would would start to play a role at much higher energies (above $\sim 100 \, \mathrm{TeV}$). Besides, spallation and nuclear decay only modify the CR spectra at low energy (below $\sim 1 \, \mathrm{GeV}$).

Therefore, from the same Green function used for the leptons, Equation \eqref{eq:solution_transport_equation_implicit}, we can reduce to the hadronic distribution function. Indeed, considering the losses as negligible, $b(E_{\mathrm{t}}) \approx b(E)$. Besides, the diffusive distance $\sqrt{r^2_{\mathrm{diff}}}$ is not dominated by the loss timescale and becomes $r^2_{\mathrm{diff}} (E) = 4 D(E) (t - t_{\mathrm{rel}})$.

In conclusion, the Green function for protons can be written as follows:
\begin{equation}\label{eq:green_function_protons}
\begin{aligned}
    f(r,t,E) &= \frac{Q(E_{\mathrm{t}})}{\pi^{3/2} r^3_{\textrm{diff}}} \cdot e^{-\frac{r^2}{r^2_{\textrm{diff}}}} \\
    &= \frac{Q(E_{\mathrm{t}})}{\left[ 4 \pi D(E) (t - t_{\mathrm{rel}}) \right]^{3/2}} \cdot e^{-\frac{r^2}{4 D(E) (t - t_{\mathrm{rel}}) }}.
\end{aligned}
\end{equation}

In the above expression we can directly implement the effect of a variable diffusion slope as $D(E) = D_0 \left( \frac{E}{E_0} \right)^{\delta(E)}$.

Finally, as done for the leptons, we get the propagated spectra integrating the Green function \eqref{eq:green_function_protons} over time, from the release time to the current instant.
\begin{figure}[ht]
    \centering
    \includegraphics[scale=0.28]{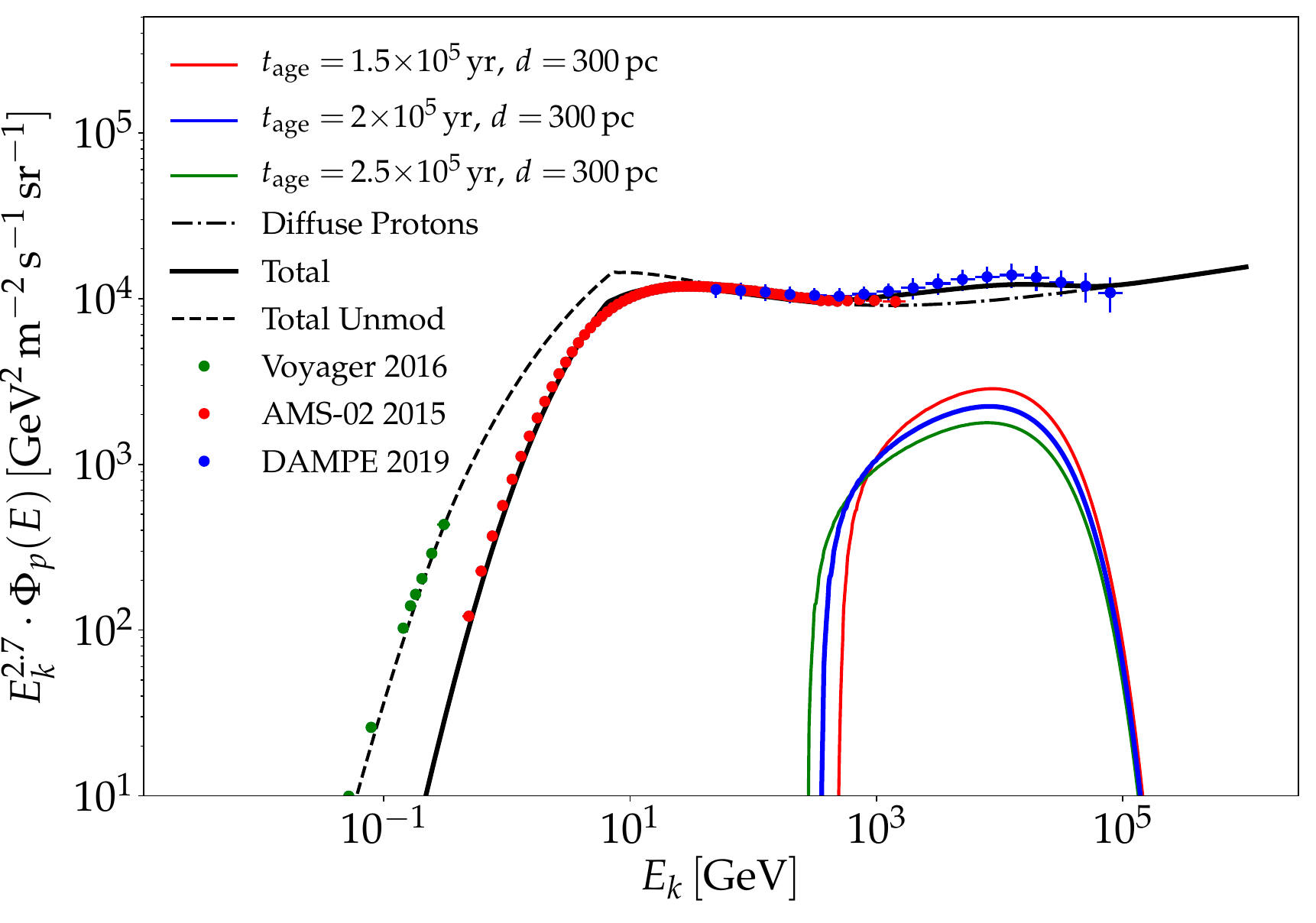}
    \caption{\small{The total proton spectrum (black solid line), resulting from the sum of the {\tt DRAGON} modulated spectrum (dashed-dotted line) and the solution of the single-source transport equation computed in this work, for the age $t_{\mathrm{age}} = 2 \cdot 10^5 \, \mathrm{yr}$ (blue solid line). Other ages (red and green solid lines), as well as the unmodulated spectrum (black dashed line) are added for comparison.}}
    \label{fig:protons_flux}
\end{figure}

In Figure \ref{fig:protons_flux}, we show our result. This is the sum of two different components:
\begin{enumerate*}[label=(\roman*)]
    \item\label{item:DRAGON_protons} the first is the diffuse CR background, \textit{i.e.} a proton population injected with slope $\Gamma^{{\tt DRA} \, p}_{\mathrm{inj}} \simeq 2.4$ and propagated with {\tt DRAGON} as described in Section \ref{sec:our_transport_setup}. This component is shown unmodulated (dashed line) and modulated with the average effective potential discussed above $\left\langle \phi_{\mathrm{mod}} \right\rangle = 0.54$ (dashed-dotted line);
    \item\label{item:nearby_source_protons} the single source, namely an injection spectrum $S(E)$ parametrized as a power-law with slope $\Gamma_{\mathrm{inj}}^{p} = 2.1$, plus a data-driven high-energy exponential cutoff implemented at $E_{\mathrm{cut}} = 20 \, \mathrm{TeV}$: we notice that such CR escape-energy from a $\sim 10^5 \, \mathrm{yr}$ source is compatible with a maximum escape energy of $1$ to a few $\mathrm{PeV}$ considered at the beginning of its \textit{Sedov} phase, namely at $\sim 10^3 \, \mathrm{yr}$. This contribution is computed by solving Equation \eqref{eq:transport_equation_polar_coordinates}, in the limit of negligible losses ($b(E_{\mathrm{t}}) \approx b(E) \rightarrow 0$), and shown for three different ages.
\end{enumerate*}

The total energy budget of the proton population originating in the source is calculated again via Equation \eqref{eq:total_energy_budget} and found to be $E^{p}_{\mathrm{tot}} \simeq 2.5 \cdot 10^{49} \, \mathrm{erg}$ for the source of our choice, \textit{i.e.} the one of age $t_{\mathrm{age}} = 2 \cdot 10^5 \, \mathrm{yr}$. The model is plotted against data points from AMS-02~\citep{Aguilar:2015ooa} and DAMPE~\citep{An:2019wcw} in the whole energy range. Furthermore, Voyager data~\citep{Cummings_2016} are also reported in the plot and appear consistent with the unmodulated propagated spectrum. Finally, the modulated sum of the two contributions is shown as the black solid line. As for the case of the all-lepton spectrum, the effect of the energy-dependent release time cuts off the low-energy ($E \lesssim 100 \, \mathrm{GeV}$) part of the spectrum. We checked that varying the parameters of the luminosity function $(\alpha_d, \tau_d)$ in the same range considered for the leptons has no sizeable impact on the proton flux.

Even though the nearby-source contribution is small, we want remark its importance for two main reasons:
\begin{enumerate}
    \item as we easily notice, without it the DAMPE points could not be reproduced,
    \item it must be present, since, as remarked before, Supernova Remnants inject both electrons and protons.
\end{enumerate}
In particular, the last statement is supported by what we find in terms of the two populations' energy budgets. In fact, the factor $E^{e^-}_{\mathrm{tot}} / E^{p}_{\mathrm{tot}} \simeq 1 \%$, as well as the two quantities evaluated separately, are consistent with the theoretical predictions of a total energy budget for a SN explosion of $E_{\mathrm{SNR}} \sim 10^{51} \, \mathrm{erg}$, a conversion efficiency in protons of the order $\sim 10^{-1} - 10^{-2}$, and in electrons of the order $\sim 10^{-3} - 10^{-5}$~\citep{Tatischeff:2009kh, Bell:2013vxa, Zirakashvili2017}.

\subsection{CR dipole anisotropy}\label{sec:CR_dipole_anisotropy}

The cosmic-ray dipole anisotropy (DA) provides a crucial complementary probe that makes it possible to constrain the model proposed in this paper. The high degree of isotropy (up to 1 part in $\sim 10^3$) detected by a variety of experiments in a wide energy range is especially constraining as far as the contribution from a local source is concerned. In particular, the interpretation of a single source as being at the origin of the spectral feature in the proton spectrum between 1 TeV and 10 TeV is heavily challenged in the context of a simple diffusion setup characterized by a single power-law. This consideration led the authors of several recent papers to consider more complex diffusion scenarios featuring an extended high-confinement zone surrounding the source of interest (see for instance \citet{Fang:2020cru}).

In this section, we consider instead the transport scenario suggested by the hardening in the light nuclei, as described in Section \ref{sec:our_transport_setup}, and compute the dipole anisotropy associated with the hidden remnant, with the formalism described below.

The CR dipole anisotropy is the first order of the expansion in spherical harmonics of the CR intensity as a function of the arrival direction, $I(\theta, \phi)$~\citep{Ahlers:2016rox}. In the case of an isolated nearby source, the dipole term is dominant and can be written as follows~\citep{Ginzburg1964}:
\begin{equation}\label{eq:dipole_anisotropy_amplitude}
    I(\alpha) = \bar{I} + \delta_i \bar{I} \, \cos \alpha, \qquad  \delta_i = \frac{I_{\mathrm{max}} - I_{\mathrm{min}}}{I_{\mathrm{max}} + I_{\mathrm{min}}},
\end{equation}
where $\alpha$ is the angle of the observation line, denoted as $\hat{n}$, with respect to the source direction, labelled $\hat{r}$.

In the diffusive-regime approximation, we obtain:
\begin{equation}\label{eq:dipole_anisotropy_diffusive}
    \delta_i = \frac{3 D(E)}{c} \Bigg| \frac{{\nabla} f_i}{f_i} \Bigg|,
\end{equation}
where $f_i \equiv f_i(r, t, E)$ is the distribution function of the cosmic rays transported from the single source.

The total dipole anisotropy, assuming the presence of a set of sources, can be written as:
\begin{equation}\label{eq:total_anisotropy}
    \Delta_{\mathrm{tot}} = \frac{\sum_i f_i \, \delta_i \, \hat{r} \cdot \hat{n}}{\sum_i f_i}.
\end{equation}

If we directly observe in the direction of the anisotropy source, $\hat{r} \cdot \hat{n} = 1$, and the total anisotropy can be decomposed as the part coming from the dominant source plus an average term coming from the background:
\begin{equation}\label{eq:anisotropy_decomposed}
    \Delta_{\mathrm{tot}} \simeq \frac{f_i \, \delta_i}{\sum_i f_i} + \biggl< \frac{\sum_i f_i \, \delta_i}{\sum_i f_i} \biggr>.
\end{equation}

To support the interpretation of the total anisotropy as two separate terms, we notice that, at the energy where the anisotropy amplitude presents an evident break ($E \sim 100 \, \mathrm{GeV}$), we also observe phase flip from R.A.$\simeq 4h$ to the direction of the Galactic Center (GC) (see \citet{Ahlers:2016rox}, their Figure 7). In other words, the DA data above this energy can be associated to the large-scale diffuse background and are assumed to follow a simple power-law~\citep{Ahlers:2016rox}. It is worth mentioning that the anisotropy associated to the diffuse cosmic rays, in principle, should directly come from the propagated distribution function computed with {\tt DRAGON}. However, we propagated the particles with a homogeneous diffusion coefficient, neglecting the vertical component of the Galactic magnetic field in the GC region. In terms of the associated $\gamma$-rays, this simplification leads to what is referred to as the \textit{gradient problem}, \textit{i.e.} the well known discrepancy (for $E_{\gamma} \geq 100 \, \mathrm{MeV}$) between the theoretical CR-flux profile obtained by assuming SNRs to be the sources of Galactic CRs and that inferred from EGRET $\gamma$-ray diffuse observations~\citep{Hunger:1997we}. Physically speaking, ignoring the vertical escape of CRs around the GC causes a longer residence time (\textit{i.e.} less-efficient diffusion) --- with respect to the exact $D(E)$ parametrization --- of the particles around the Galactic Center, resulting in a larger production of photons. Analogously, we would expect the same overproduction of CRs in the GC region to overestimate the real dipole anisotropy.

\begin{figure}[ht]
    \centering
    \includegraphics[scale=0.28]{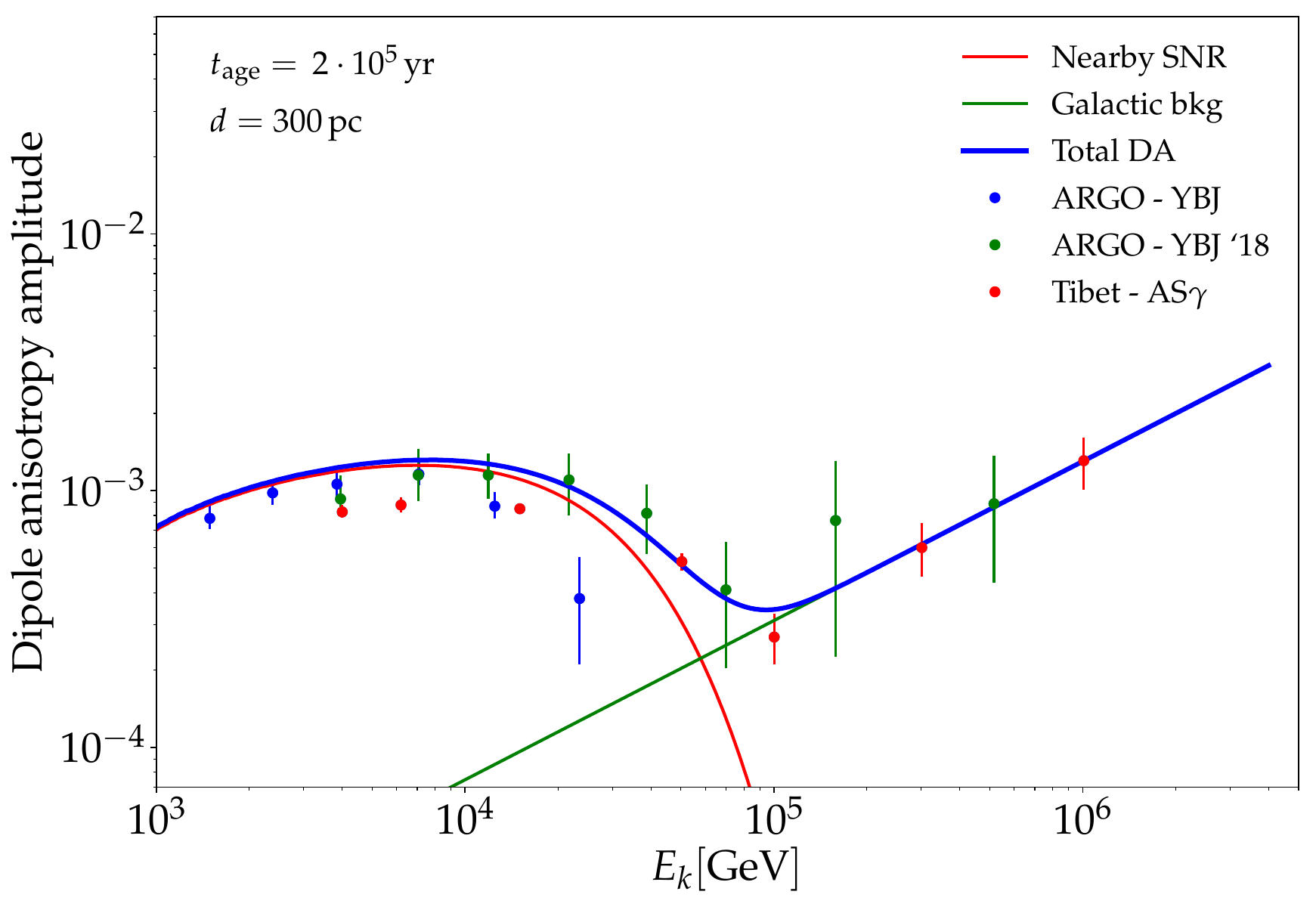}
    \caption{\small{Cosmic-ray dipole anisotropy amplitude calculated as the sum of a background anisotropy (green solid line) and the single source contribution (red solid line) for the source of age $t_{\mathrm{age}} = 2 \cdot 10^5 \, \mathrm{yr}$. Anisotropy data are consistent with each other, therefore here we plot a subset of them, to avoid confusion. The plotted points are from ARGO~\citep{Bartoli:2015ysa,Bartoli:2018ach} and Tibet-AS$\gamma$~\citep{Amenomori:2017jbv}.}}
    \label{fig:CR_dipole_anisotropy}
\end{figure}

Motivated by these considerations, in Figure \ref{fig:CR_dipole_anisotropy} we show that the hypothesis of one nearby old remnant originating the CR populations, responsible for both the leptonic and the hadronic features, is compatible with the current anisotropy data.

To reproduce the diffuse contribution, we use the fit parameters recently suggested in \citet{Fang:2020cru}, according to which the background anisotropy can be written as $\Delta_{\mathrm{bkg}} = c_1 \left( \frac{E}{1 \, \mathrm{PeV}} \right)^{c_2}$, where $(c_1, \, c_2) = (1.32 \cdot 10^{-3}, \, 0.62)$. The result is the green solid line in the figure. 

On the other hand, the single-source contribution is found under the assumption of diffusive behaviour for the released particles. This component corresponds to the red solid line in the figure, for the source of age $t_{\mathrm{age}} = 2 \cdot 10^5 \, \mathrm{yr}$, the same considered in the previous sections.

We want to remark again that a key role to reproduce the observations is here played by the slope of the diffusion coefficient, that, according to Equation \eqref{eq:delta_tomassetti}, becomes harder in the high-energy region ($\delta \lesssim 0.2$ at $E > 10 \, \mathrm{TeV}$).

\section{Discussion}\label{sec:discussion}

As a first discussion point, we want to comment on the nature of the source here invoked. Given its old age, it is reasonable to assume that the remnant is currently in the final stage of its evolution, deep into the radiative phase. Hence, we expect it to be quite extended and the detection of its faint multi-wavelength signature to be very challenging, especially from a distance as large as $\sim 300 \, \mathrm{pc}$. In particular, if $\sim 100 \, \mathrm{GeV}$ protons are still confined in the SNR at its age, then one should expect a $\gamma$-ray emission, resulting from pion decay, cutting off around $\sim 10 \, \mathrm{GeV}$. Electrons at these energies emit synchrotron radiation up to a frequency of $\sim 300 \, \mathrm{GHz}$ and the source may be of interest for future {\it Square-Kilometer Array} (SKA) observations. Moreover, electrons contribute to IC $\gamma$-ray emission up to $\sim 100 \, \mathrm{MeV}$, $\sim 1 \, \mathrm{GeV}$ and $\sim 10 \, \mathrm{GeV}$ for IR, optical, UV soft photons background.

From the point of view of the injection mechanisms, we notice that the injection slopes of the protons and leptons from our additional source are different from the ones considered for the large-scale background computed with {\tt DRAGON}: this is due to the fact that the latter component results from the convolution of a large number of sources. Therefore, the information on their possibly different injections and the locations of their emission peaks is averaged out. However, as a consistency check, we observe that the discrepancy between proton's and lepton's spectral indices is the same if we consider separately the single source ($\Delta \Gamma_{\mathrm{inj}} = \Gamma_{\mathrm{inj}}^{e^-} - \Gamma_{\mathrm{inj}}^{p} = 0.35$) and the large-scale distribution ($\Delta \Gamma_{\mathrm{inj}}^{\tt DRA} = \Gamma_{\mathrm{inj}}^{{\tt DRA} \, e^-} - \Gamma_{\mathrm{inj}}^{{\tt DRA} \, p} = 0.34$). This in fact well reflects the different physics behind proton's and lepton's release, as discussed at the end of Section \ref{subsec:all_lepton_spectrum}.

From a wider prospective, regarding a SNR origin for the $e^{+} + e^{-}$ spectrum, it is worth noticing that, due to the incompleteness of the catalogues, especially for old remnants, a proof of concept would be represented by a Monte Carlo simulation of all the possible configurations of source distributions in our Galaxy. A step in this direction is presented in \citet{2020arXiv201011955E}, suggesting that the SNR explanation is disfavored at more than $2\sigma$, with respect to the average configuration. This result is model dependent, in particular is based on a source distribution that is set to follow the Galactic spiral arms. However, the Solar system is found in the so-called \textit{Orion Spur}, a minor arm-structure in the Milky Way between two major arms. This is not included in that work, whereas we believe it to be of major importance, in particular for the leptonic observables above $1 \, \mathrm{TeV}$. The amount of the uncertainty can be estimated in their Figure 6, where the lepton horizons --- as caused by their energy-loss rate --- for particles of $E = 100 \, \mathrm{GeV}, \, 1 \, \mathrm{TeV}, 10 \, \mathrm{TeV},$ are sketched. In particular, the $10 \, \mathrm{TeV}$ horizon includes two arcs of two major arms at equal distance from the Solar system. Therefore, one estimates that ignoring the \textit{Orion Spur} results in neglecting roughly $\sim 1/3$ of the leptons of this energy. Similarly, we estimate that $\sim 20/25\%$ of the particles are missing from the $1 \, \mathrm{TeV}$ range of the $e^+ + e^-$ spectrum. This lack can abundantly account for the $2\sigma$ dispersion of the Monte Carlo average curve. It is therefore the reason why, on average, the high-energy ($E \sim 1 \, \mathrm{TeV}$) range of the all-lepton spectrum cannot be captured by their calculations.

In this context, we want to comment on the number of nearby SNRs that we may expect to contribute to the high-energy part of the leptonic and hadronic spectra. We remark that a limited number of young sources exist in the vicinity of the Sun, and they may also provide a sizable contribution to the observed fluxes. In particular, we emphasize the possible role of the young type II Supernova Remnant in the southern constellation Vela. The young age of this accelerator ($\simeq 1.1 \cdot 10^4 \, \mathrm{yr}$) restricts its potential signature in the lepton spectrum at energies as large as $\sim 10^4\, \mathrm{GeV}$, thus not limiting our proposed scenario. However, its presence could constrain the parameters involved in the luminosity function and in the energy-dependent release time. Indeed, a rough calculation of its emission based on our reference transport setup has revealed a predicted flux that is strongly dependent on the parameters of the model, and that can span between a negligible contribution --- as small as more than 2 orders of magnitude below the level of the data points --- and a dominant one. However, a detailed modeling of this object constrained by multi-wavelength data is beyond the scope of the present work.

Moreover, we are confident that more accurate data in this domain --- $E \sim 1 - 50 \, \mathrm{TeV}$, subject of interest for the \textit{Cherenkov Telescope Array} (CTA) --- expected in the near future will help to disentangle the question, possibly revealing the presence of a spectral feature.

A final important point that is worth to discuss regards the implications of using the same rigidity-dependent diffusion coefficient for both the diffuse CR component and the isolated nearby source. In particular, this means that hardening at $\sim 200 \, \mathrm{GeV}$ is actually due to a superposition of two effects:
\begin{enumerate*}[label=(\roman*)]
    \item\label{item:diffusive_hardening} the diffusive origin coming from physical differences in the halo and in the disk;
    \item\label{item:nearby_source} the nearby-source contribution.
\end{enumerate*} In this sense, an important role is played by the softening in the DAMPE spectrum, that is interpreted as an intrinsic cutoff of the hidden remnant. In fact, even though a more pronounced hardening with no additional sources could be considered to account for the mismatch between AMS-02 and DAMPE data, this would be still not sufficient to reproduce the complex structure observed by DAMPE --- the \textit{softening} at $E \sim 10 \, \mathrm{TeV}$. In particular, no theoretical models predict so far a cutoff in the proton propagated spectra below the \textit{knee} ($E_{\mathrm{knee}} \sim 5 \, \mathrm{PeV}$). As a consequence, the scenario here proposed predicts that the CR spectrum above $E \sim 100 \, \mathrm{TeV}$ would have a slope similar to that observed after the $\sim 200 \, \mathrm{GeV}$ hardening. With this aim, higher energy data points in the future will certainly help to disentangle this puzzle.

\section{Conclusions}\label{sec:conclusions}

In this paper we proposed the idea that the spectral feature at $\sim 10 \, \mathrm{TeV}$ in the cosmic-ray proton spectrum recently reported by the DAMPE Collaboration together with the spectral break at $\sim 1 \, \mathrm{TeV}$ measured by  H.E.S.S. in the lepton spectrum have a common origin and can be associated to a nearby, fading Supernova Remnant. We believe this simultaneous interpretation to be of paramount importance, since SNRs are accelerators for both electrons and protons.

We injected the particles with a realistic --- and physically motivated --- energy-dependent release time that considers the different stages of the SNR evolution and the surrounding medium, and with a luminosity function that declines over time. Then, we computed their propagation from such object in a spherically symmetric setup, and found that all the available observables can be simultaneously reproduced.  
The key ingredient in the calculation is a transport setup based on a diffusion coefficient characterized by a smooth transition to a progressively harder rigidity-scaling at higher energies, as suggested by the light nuclei spectra measured by the AMS-02 Collaboration. This feature allowed us to reproduce the cosmic-ray anisotropy data without any further assumption. Moreover, the combined leptonic and hadronic data led us to characterize the properties of the particles accelerated by such object in good agreement with theoretical expectations.

%The combined leptonic and hadronic data allowed us to characterize the properties of the particles accelerated by such object: in particular, we obtained  a power-law proton injection slope $\Gamma_{\mathrm{inj}}^{p} = 2.1$
%and an electron injection slope $\Gamma_{\mathrm{inj}}^{e^-} = 2.4$, in good agreement with theoretical expectations.  %Finally, we found that $\mathcal{O}(10\%)$ of the source energy budget goes into acceleration of hadrons, and $\mathcal{O}(1\%)$ is converted into accelerated leptons.

\section*{Acknowledgements}
We are grateful to the anonymous referees for the many suggestions, they helped a lot improving the clarity of the manuscript. Also, we thank C. Evoli, S. Gabici, D. Grasso, P.S. Marrocchesi for many inspiring discussions and for useful comments on this work.

O.F. was supported by the University of Siena with a joint doctoral degree with the Autonomous University of Madrid. 

D.G.\ has received financial support through the Postdoctoral Junior Leader Fellowship Programme from la Caixa Banking Foundation (grant n.~LCF/BQ/LI18/11630014).
D.G. was also supported by the Spanish Agencia Estatal de Investigaci\'{o}n through the grants PGC2018-095161-B-I00, IFT Centro de Excelencia Severo Ochoa SEV-2016-0597, and Red Consolider MultiDark FPA2017-90566-REDC.

\clearpage
%\appendix
\begin{appendices}
\numberwithin{equation}{section}

\section{Energy-dependent release time from Supernova remnant shocks}\label{app:release_time}
In this appendix we review the dominant mechanisms that confine particles inside the Supernova shocks. Once those processes are overcome, particles can be released from the source. As leptons suffer from severe energy losses and are $m_p/m_e \sim 10^3$ times less efficient than hadrons in generating streaming instabilities, the release processes for hadrons and leptons will be discussed separately.

\subsubsection{Release time for hadrons}
Hadrons can escape from SNRs because of two main reasons: 
\begin{enumerate*}[label=(\roman*)]
    \item\label{item:geometrical_losses} due to geometrical losses, when their mean free path gets larger than a fraction of the shock radius~\citep{berezhko94};
    \item\label{item:limited_cuurent} due to the limited current they are able to trigger \textit{upstream}\footnote{The region \textit{upstream} --- as opposed to the \textit{downstream} --- of the shock is the region where the shock front has already passed.} of the shock~\citep{schure13}.
\end{enumerate*}
In the latter case, the CR current is necessary to trigger the non-resonant streaming instability and to produce magnetic field amplification at the shock precursor~\citep{bell04}. As the non-resonant instability growth rate scales as $\sim u_{\rm sh}^3$, with $u_{\mathrm{sh}}$ velocity of the shock --- for a $\propto E^{-2}$ particle distribution that we assume hereafter --- it likely controls the maximum CR energy at the early stages of the evolution of the SNR shock, {\it i.e.} during free expansion and, possibly, Sedov-Taylor phases.

Maximum energies imposed by geometrical losses are set because the CR diffusive path in the precursor reaches a fraction $\xi < 1$ of the shock radius $R_{\rm sh}$, namely 
\begin{equation}
    \ell = {D(E) \over u_{\rm sh}(t)} = \xi R_{\rm sh}(t),
\end{equation}
where the diffusion coefficient is here parametrized in terms of its Bohm value $D(E)=\eta_{\rm acc} r_{\rm L} c/3$, where $\eta_{\mathrm{acc}}$ is a numerical factor $\eta_{\rm acc} \ge 1$. We consider relativistic particles of charge $Ze$, with a Larmor radius $r_{\rm L} =E/Ze B(t)$ (hereafter we only consider protons, so $Z=1$). Therefore the maximum energy fixed by geometrical losses is
\begin{equation}
    E_{\mathrm{max, Geo}}= {3\xi e \over \eta_{\rm acc} c} R_{\rm sh}(t) u_{\rm sh}(t) B(t) . 
\end{equation}
Hereafter we fix $\xi=0.3$ and $\eta_{\rm acc}=1$.

\vspace{0.3cm}
Limited-current loss process dominates in case of strong magnetic field amplification, hence during the SNR evolution stages where the shock strength is high. The maximum CR energy in that case depends on the type of ambient medium: either circum-stellar gas (CSM) --- as for a core-collapse Supernova --- or interstellar gas (ISM) --- as for a type Ia Supernova --- \citep{schure13}:
\begin{align}
&\phi E_{\rm esc, Cur, CSM} = {e \sqrt{\pi} \over \gamma \tau c} \chi u_{\rm sh}(t)^2 R_{\rm sh}(t) \sqrt{\rho(t)} \ ,\label{eq:escape_protons_CSM} \\
&\phi E_{\rm esc, Cur, ISM} = {e \sqrt{\pi} \over 2 \gamma \tau c} \chi u_{\rm sh}(t)^2 R_{\rm sh}(t) \sqrt{\rho},
\end{align}
where $\gamma \tau$ is the number of e-folding growth time necessary to amplify the magnetic field (we take $\gamma \tau=5$ hereafter), $\chi= U_{\rm CR}/\rho u_{\rm sh}^2$, is the fraction of the shock kinetic energy imparted into CRs (we take $\chi=0.1$ hereafter), $\rho$ is the ambient gas mass density and $\phi =\ln(E_{p, \mathrm{max}}/m_{p}c^2)$.

We consider a shock radius scaling with time as $\sim t^b$, where $b$ depends on the evolution stage: $b=1$, $b=2/5$, $b=3/10$, $b=1/4$ in the \textit{free expansion} (Free), \textit{Sedov-Taylor} (ST), \textit{pressure-driven snowplough} (PDS) and \textit{momentum-conservation phases} (MCS), respectively. We use the scaling laws derived in \citet{truelove99,cioffi88} to evaluate the shock radius and speed at the transition between two phases. The magnetic field strength is assumed to vary as a certain power of the shock speed, namely $B(t) \propto u_{\rm sh}^a$, where $a$ may depend on the SNR evolution stage. Once the time dependence of $E_{p, \mathrm{max}}$ is explicit, we can inverse it to find the release time $t(E_{p, \mathrm{max}})$.

With this procedure, the timescales for the different stages of the SNR evolution, from the ST phase until the dissipation of the remnant (\textit{merging stage}), can be calculated as follows:
\begin{equation}\label{eq:timescales_SNR_STAGES}
\begin{aligned}
   &t_{\mathrm{ST, kyr}}  =  0.3 \, E_\mathrm{SNR,51}^{-1/2} \, M_{\mathrm{ej},\odot} \, n_{T,1}^{-1/3} \\
   &t_{\mathrm{PDS, kyr}}  =  \frac{36.1 \, e^{-1} \, E_\mathrm{SNR,51}^{3/14}}{\xi_n^{5/14} \, n_{T,1}^{4/7}} \\
   &t_{\mathrm{MCS, kyr}}  =  \mathrm{min} \left[ \frac{61 \, v_\mathrm{ej,8}^3}{\xi_n^{9/14} \, n_{T,1}^{3/7} \, E_\mathrm{SNR,51}^{3/14}}, \frac{476}{(\xi_n \Phi_c)^{9/14}} \right] t_\mathrm{PDS, kyr} \\ &t_{\mathrm{merge, kyr}} = 153 \left(\frac{E_\mathrm{SNR,51}^{1/14} \, n_{T,1}^{1/7} \, \xi_n^{3/14}}{\beta \, C_{06}} \right)^{10/7} t_\mathrm{MCS, kyr},
\end{aligned}
\end{equation}
where $E_{\mathrm{SNR, 51}}$ is the total energy of the SN explosion in units of $10^{51} \, \mathrm{erg}$, $M_{\mathrm{ej}, \odot}$ is the mass of the ejected material in units of $1$ Solar masses, $n_{T, 1} = \rho/m_p$ is the ambient medium density in units of $1 \, \mathrm{cm^{-3}}$, $\xi_n$ is the ambient medium metallicity, $v_{\mathrm{ej},8}$ is the speed of the ejected material in units of $10^8 \, \mathrm{cm / s}$, $\Phi_c = 1$ is the thermal plasma conductivity, $\beta = 2$ is the factor by which the pressure inside the shock exceeds the ambient-medium pressure and $C_{06} = 1$ is the sound speed in units of $10^6 \, \mathrm{cm/s}$. In this work, we fix the energy budget to $E_{\mathrm{tot, SNR}} = 10^{51} \, \mathrm{erg}$, the ejecta mass $M_{\mathrm{ej}} = 1 \, M_\odot$, the ejecta velocity to $v_{\mathrm{ej}} = 10^9 \, \mathrm{cm/s}$ and the ambient density to $n_{T}=10~\rm{cm^{-3}}$. These timescales are expressed in kiloyears.

\begin{table*}[!t]
    \centering
    \def\arraystretch{1.5} % rows separation
    \setlength\tabcolsep{0.5cm} % columns separation
    \begin{tabular}{ c|c|c|c }
         & Chosen value & Range & Comments \& References \\
         \hline
         $\xi$ & 0.3 & $[0.05, \, 0.3]$ & \cite{1996APh.....5..367B} \\
         $\gamma \tau$ & 5 & $[3, \, 9]$ & \cite{schure13} \\
         $\chi$ & 0.1 & $[0.1, \, 0.5]$ & \cite{bell04} \\
         $\eta_{\mathrm{acc}}$ & 1 & --- & Bohm diffusion for efficient DSA \\
         $n_{T}$ & $10 \, \mathrm{cm^{-3}}$ & $[0.01, \, 10^{2}]$ & based on the region of the ISM \\
         $\xi_n$ & 1 & --- & \cite{cioffi88} \\
         $\beta$ & 2 & $[1, \, 3]$ & \cite{cioffi88} \\
         $C_{06}$ & 1 & $[1, \, 10]$ & \cite{cioffi88} \\
         $v_{\mathrm{ej}}$ & $10^9 \, \mathrm{cm/s}$ & --- & for a SN of $1 M_{\odot}$ and $10^{51} \, \mathrm{erg}$ \\
         $\Phi_c$ & 1 & --- & \cite{1982ASIC...90..433M}
    \end{tabular}
    \caption{\small{The table reports a list of the parameters that regulate the particle release in the ISM. The first three rows control the maximum energy that hadrons can have to be trapped within the shock, while the others control the timescales of the evolution stages of the SNR, hence applying to both hadrons and leptons.}}
    \label{tab:my_label}
\end{table*}

Table \ref{tab:my_label} reports a summary of such parameters and shows that they are well within accepted uncertainty ranges. We verified that our findings are robust with respect to the uncertainty intervals, with only the ISM density $n_{T}$ producing a small but sizeable effect on the nearby-source solution. In particular, as it can be seen from Equation \eqref{eq:timescales_SNR_STAGES}, changing $n_{T}$ affects the SNR evolution-timescales and, as a consequence, the CR release time. However, as already mentioned in the introduction to Section \ref{sec:nearby_source_contributions}, the age of the particle is estimated as $t_{\mathrm{CR}, \mathrm{age}} = t_{\mathrm{rel}} + \Delta t_{\mathrm{travel}}$ and, according to \eqref{eq:timescales_SNR_STAGES} and to the required escape energy as a function of time --- $E_{\mathrm{esc}} \sim t^{-11/10}$, $E_{\mathrm{esc}} \sim t^{-6/5}$ or $E_{\mathrm{esc}} \sim t^{-5/4}$ ---, above $\sim 1 \, \mathrm{TeV}$ we are observing particles whose age is dominated by the diffusive time, while below $\sim \mathcal{O}(100) \, \mathrm{GeV}$ their age is dominated by the release time. Therefore, varying $n_T$ within the uncertainty range will have an impact only in the low-energy part of the spectrum. Moreover, the change is sufficiently small that can be easily reabsorbed either changing the injection parameters of the large-scale CR-distribution or with a further fine tuning of the transport setup found in \citep{Tomassetti:2012ga}.

In this work we consider that the maximum CR energy is current-limited in the free expansion and Sedov phases, while it is limited by geometrical losses during the later radiative phases. Strong magnetic field amplification only occurs during the first two adiabatic phases. The magnetic field is assumed to scale as $u_{\rm sh}^{3/2}$ in the adiabatic phases and as $u_{\rm sh}$ in the radiative phases (see discussion in \citet{voelk05}). We further assume that the maximum magnetic field strength and the maximum CR energy are reached at the start of the Sedov phase. They are fixed to $100 \, \mu \mathrm{G}$ and $1 \, \mathrm{PeV}$ respectively.

To summarize, we used Equation \eqref{eq:escape_protons_CSM} to calculate the proton escape energy as a function of time as follows:
\begin{align*}
    &\begin{aligned}
        \bullet \; &\ln \left( \frac{E_{\mathrm{esc, Cur}} (t)}{m_p c^2} \right) E_{\mathrm{esc, Cur}}(t) = \ln(E_M (t_{\mathrm{ST}})) \left( \frac{t}{t_{\mathrm{ST}}} \right)^{-6/5} \\
        &\text{such that } E_M \equiv E_{p, \mathrm{max}}(t_{\mathrm{ST}}) = 1 \, \mathrm{PeV}
    \end{aligned}\\[8pt]
    &\begin{aligned}
        \bullet \; E_{\mathrm{esc, Geo, 1}}(t) &= E_{\mathrm{M}} (t_{\mathrm{PDS}}) \left( \frac{t}{t_{\mathrm{PDS}}} \right)^{-11/10} \\
        &= E_{\mathrm{esc, Cur}} (t_{\mathrm{PDS}}) \left( \frac{t}{t_{\mathrm{PDS}}} \right)^{-11/10} 
    \end{aligned}\\[8pt]
    &\begin{aligned}
        \bullet \; E_{\mathrm{esc, Geo, 2}}(t) &= E_{\mathrm{M}} (t_{\mathrm{MCS}}) \left( \frac{t}{t_{\mathrm{MCS}}} \right)^{-5/4} \\
        &= E_{\mathrm{esc, Geo, 1}} (t_{\mathrm{MCS}}) \left( \frac{t}{t_{\mathrm{MCS}}} \right)^{-5/4}.
    \end{aligned}
\end{align*}

\subsubsection{Release time for leptons} Besides the processes already discussed for hadrons, leptons are also sensitive to radiative losses. The maximum energy fixed by radiative losses is $E_{e, \mathrm{max, loss}}$. These losses can prevent them to escape the SNR until the condition $E_{e, \mathrm{max, loss}} \le E_{p, \mathrm{max}}$ is fulfilled~\citep{ohira12}. The energy $E_{e, \mathrm{max}}$ is set by the condition $t_{\rm acc}= t_{\rm loss}$ where $t_{\rm acc}$ and $t_{\rm loss}$ are the acceleration and loss timescales respectively. We assume here a simple form of the acceleration timescale, $t_{\rm acc} = \eta_{\rm acc} f(r) D_{\rm Bohm}/u_{\rm sh}^2$, where $f(r)$ is a function of the shock compression ratio. For a parallel shock $f(r) \sim 3r(r+1)/(r-1)$, while, if magnetic field amplification occurs upstream of the shock, the function assumes the form $f(r) \sim 6.6r/(r-1)$~\citep{parizot06}. A compression ratio $r=4$ is adopted hereafter. The time dependence of radiative losses is imposed by the time variation of the magnetic field strength $B(t)$ in the synchrotron process. Synchrotron loss-timescale for an electron of energy $E$ is $t_{\rm loss, syn}= 6\pi m_{e}^2 c^4/ \sigma_T c B(t)^2 E$, where $m_{e}$ is the electron mass and $\sigma_{T}$ is the Thomson cross section.

In conclusion, assuming that geometrical losses are responsible for electron escape at each stage of the SN evolution from the Sedov phase on, to calculate the electron escape energy as a function of time we proceed with the following steps:

\begin{align*}
    &\begin{aligned}
        \bullet \; &E_{\mathrm{esc, Geo, 0}}(t) = E_{\mathrm{M}} (t_{\mathrm{ST}}) \left( \frac{t}{t_{\mathrm{ST}}} \right)^{-11/10}\\
        &\text{such that } E_M \equiv E_{e, \mathrm{max}}(t_{\mathrm{ST}}) = 100 \, \mathrm{TeV}
    \end{aligned} \\[8pt]
    &\begin{aligned}
        \bullet \; E_{\mathrm{esc, Geo, 1}}(t) &= E_{\mathrm{M}} (t_{\mathrm{PDS}}) \left( \frac{t}{t_{\mathrm{PDS}}} \right)^{-11/10} \\
        &= E_{\mathrm{esc, Geo, 0}} (t_{\mathrm{PDS}}) \left( \frac{t}{t_{\mathrm{PDS}}} \right)^{-11/10}
    \end{aligned} \\[8pt]
    &\begin{aligned}
        \bullet \; E_{\mathrm{esc, Geo, 2}}(t) &= E_{\mathrm{M}} (t_{\mathrm{MCS}}) \left( \frac{t}{t_{\mathrm{MCS}}} \right)^{-5/4} \\
        &= E_{\mathrm{esc, Geo, 1}} (t_{\mathrm{MCS}}) \left( \frac{t}{t_{\mathrm{MCS}}} \right)^{-5/4}.
    \end{aligned}
\end{align*}

\section{On the expected number of nearby hidden remnants}\label{app:numberRemnants}

In this appendix, we discuss the motivations to consider only one additional source to look for in the vicinity of the Earth. We consider the rate --- per unit volume, at the solar circle, as a function of the Galactic latitude $z$ --- of both type Ia and type II Supernova events that are implemented in {\tt DRAGON}~\citep{Ferriere:2001rg}:

\begin{equation}\label{eq:supernova_rate_ferriere}
\begin{aligned}
    &\mathcal{R_{\mathrm{I}}} (z) = \left( 7.3 \, \mathrm{kpc^{-3} \, Myr^{-1}} \right) \cdot  e^{-\frac{|z|}{325 \, \mathrm{pc}}} \\
    &\begin{aligned}
    \mathcal{R_{\mathrm{II}}} (z) = \left( 50 \, \mathrm{kpc^{-3} \, Myr^{-1}} \right) \cdot & \left\{ 0.79 \, e^{- \left( \frac{|z|}{212 \, \mathrm{pc}} \right)^2} \right. \\
     &\left. + \, 0.21 \, e^{- \left( \frac{|z|}{636 \, \mathrm{pc}} \right)^2}\, \right\}.
    \end{aligned}
\end{aligned}
\end{equation}

Since we are testing the hypothesis of a Supernova as source of high-energy leptons ($E_{e^{\pm}} > 1 \, \mathrm{TeV}$), we integrate those rates in a cylinder of half-height $h_{\mathrm{cyl}} = 1 \, \mathrm{kpc}$, as this is roughly the distance that those leptons can travel, due to their massive energy-loss. Thus we need to compute:
\begin{equation}\label{eq:supernova_events_integral}
    n_{\mathrm{SNR}} \, \mathrm{[kpc^{-2} \cdot Myr^{-1}]} = \int_{-1 \, \mathrm{kpc}}^{+1 \, \mathrm{kpc}} dz \, \left( \mathcal{R_{\mathrm{I}}} (z) + \mathcal{R_{\mathrm{II}}} (z) \right).
\end{equation}

The result of the integral has to be multiplied by the base area of the cylinder $A = \pi r_{\mathrm{cyl}}^2$, where $r_{\mathrm{cyl}} = 1 \, \mathrm{kpc}$ for the same losses reasons, and by the lifetime of a typical Supernova Remnant, $\tau_{\mathrm{age}} \sim 5 \cdot 10^5 \, \mathrm{yr}$. Therefore, within one SNR lifetime and $1 \, \mathrm{kpc}$ from the Earth, we expect $N_{\mathrm{SNR}} \simeq 2.2$ Supernova Remnants potentially contributing to the observed lepton flux. 

Since we already observe five of them~\citet{Fornieri:2019ddi}, we expect the lowest possible number of additional hidden sources to dominate the observed all-lepton spectrum on Earth. This assumption is corroborated by the observation of a directional bump in the dipole anisotropy amplitude (see \citet{Ahlers:2016rox} and references therein), as discussed in Section \ref{sec:CR_dipole_anisotropy}.

As a comment on the estimation of the event rate, it might be argued that the Solar system is embedded in what is referred to as the Local Bubble, a low-density ($n_{\mathrm{HI}} \lesssim 0.1 \, \mathrm{cm^{-3}}$) region of the Galaxy of radius $r_{\mathrm{LB}} > 300 \, \mathrm{pc}$ that likely originated by the explosion of several SNe~\citep{2020A&A...636A..17P}. This could imply a different rate of Supernova events inside it. However, since the age of the Bubble is estimated to be $\sim \mathcal{O}(10^7)$ yrs, which is much larger than the average lifetime of a SN, this can only affect the calculation in the sense of lowering the number of expected events.
\end{appendices}

\clearpage
\bibliography{apssamp}

\providecommand{\noopsort}[1]{}\providecommand{\singleletter}[1]{#1}%
\begin{thebibliography}{77}%
\makeatletter
\providecommand \@ifxundefined [1]{%
 \@ifx{#1\undefined}
}%
\providecommand \@ifnum [1]{%
 \ifnum #1\expandafter \@firstoftwo
 \else \expandafter \@secondoftwo
 \fi
}%
\providecommand \@ifx [1]{%
 \ifx #1\expandafter \@firstoftwo
 \else \expandafter \@secondoftwo
 \fi
}%
\providecommand \natexlab [1]{#1}%
\providecommand \enquote  [1]{``#1''}%
\providecommand \bibnamefont  [1]{#1}%
\providecommand \bibfnamefont [1]{#1}%
\providecommand \citenamefont [1]{#1}%
\providecommand \href@noop [0]{\@secondoftwo}%
\providecommand \href [0]{\begingroup \@sanitize@url \@href}%
\providecommand \@href[1]{\@@startlink{#1}\@@href}%
\providecommand \@@href[1]{\endgroup#1\@@endlink}%
\providecommand \@sanitize@url [0]{\catcode `\\12\catcode `\$12\catcode
  `\&12\catcode `\#12\catcode `\^12\catcode `\_12\catcode `\%12\relax}%
\providecommand \@@startlink[1]{}%
\providecommand \@@endlink[0]{}%
\providecommand \url  [0]{\begingroup\@sanitize@url \@url }%
\providecommand \@url [1]{\endgroup\@href {#1}{\urlprefix }}%
\providecommand \urlprefix  [0]{URL }%
\providecommand \Eprint [0]{\href }%
\providecommand \doibase [0]{https://doi.org/}%
\providecommand \selectlanguage [0]{\@gobble}%
\providecommand \bibinfo  [0]{\@secondoftwo}%
\providecommand \bibfield  [0]{\@secondoftwo}%
\providecommand \translation [1]{[#1]}%
\providecommand \BibitemOpen [0]{}%
\providecommand \bibitemStop [0]{}%
\providecommand \bibitemNoStop [0]{.\EOS\space}%
\providecommand \EOS [0]{\spacefactor3000\relax}%
\providecommand \BibitemShut  [1]{\csname bibitem#1\endcsname}%
\let\auto@bib@innerbib\@empty
%</preamble>
\bibitem [{\citenamefont {Gabici}\ \emph {et~al.}(2019)\citenamefont {Gabici},
  \citenamefont {Evoli}, \citenamefont {Gaggero}, \citenamefont {Lipari},
  \citenamefont {Mertsch}, \citenamefont {Orlando}, \citenamefont {Strong},\
  and\ \citenamefont {Vittino}}]{Gabici:2019jvz}%
  \BibitemOpen
  \bibfield  {author} {\bibinfo {author} {\bibfnamefont {S.}~\bibnamefont
  {Gabici}}, \bibinfo {author} {\bibfnamefont {C.}~\bibnamefont {Evoli}},
  \bibinfo {author} {\bibfnamefont {D.}~\bibnamefont {Gaggero}}, \bibinfo
  {author} {\bibfnamefont {P.}~\bibnamefont {Lipari}}, \bibinfo {author}
  {\bibfnamefont {P.}~\bibnamefont {Mertsch}}, \bibinfo {author} {\bibfnamefont
  {E.}~\bibnamefont {Orlando}}, \bibinfo {author} {\bibfnamefont
  {A.}~\bibnamefont {Strong}},\ and\ \bibinfo {author} {\bibfnamefont
  {A.}~\bibnamefont {Vittino}},\ }\bibfield  {title} {\bibinfo {title} {{The
  origin of Galactic cosmic rays: challenges to the standard paradigm}},\
  }\href {https://doi.org/10.1142/S0218271819300222} {\bibfield  {journal}
  {\bibinfo  {journal} {Int. J. Mod. Phys. D}\ }\textbf {\bibinfo {volume}
  {28}},\ \bibinfo {pages} {1930022} (\bibinfo {year} {2019})},\ \Eprint
  {https://arxiv.org/abs/1903.11584} {arXiv:1903.11584 [astro-ph.HE]}
  \BibitemShut {NoStop}%
\bibitem [{\citenamefont {{Amato}}\ and\ \citenamefont
  {{Blasi}}(2018)}]{2018AdSpR..62.2731A}%
  \BibitemOpen
  \bibfield  {author} {\bibinfo {author} {\bibfnamefont {E.}~\bibnamefont
  {{Amato}}}\ and\ \bibinfo {author} {\bibfnamefont {P.}~\bibnamefont
  {{Blasi}}},\ }\bibfield  {title} {\bibinfo {title} {{Cosmic ray transport in
  the Galaxy: A review}},\ }\href {https://doi.org/10.1016/j.asr.2017.04.019}
  {\bibfield  {journal} {\bibinfo  {journal} {Advances in Space Research}\
  }\textbf {\bibinfo {volume} {62}},\ \bibinfo {pages} {2731} (\bibinfo {year}
  {2018})},\ \Eprint {https://arxiv.org/abs/1704.05696} {arXiv:1704.05696
  [astro-ph.HE]} \BibitemShut {NoStop}%
\bibitem [{\citenamefont {Aguilar}\ \emph
  {et~al.}(2015{\natexlab{a}})\citenamefont {Aguilar} \emph
  {et~al.}}]{PhysRevLett.114.171103}%
  \BibitemOpen
  \bibfield  {author} {\bibinfo {author} {\bibfnamefont {M.}~\bibnamefont
  {Aguilar}} \emph {et~al.} (\bibinfo {collaboration} {AMS Collaboration}),\
  }\bibfield  {title} {\bibinfo {title} {{Precision Measurement of the Proton
  Flux in Primary Cosmic Rays from Rigidity 1 GV to 1.8 TV with the Alpha
  Magnetic Spectrometer on the International Space Station}},\ }\href
  {https://doi.org/10.1103/PhysRevLett.114.171103} {\bibfield  {journal}
  {\bibinfo  {journal} {Phys. Rev. Lett.}\ }\textbf {\bibinfo {volume} {114}},\
  \bibinfo {pages} {171103} (\bibinfo {year} {2015}{\natexlab{a}})}\BibitemShut
  {NoStop}%
\bibitem [{\citenamefont {Aguilar}\ \emph {et~al.}(2020)\citenamefont {Aguilar}
  \emph {et~al.}}]{PhysRevLett.124.211102}%
  \BibitemOpen
  \bibfield  {author} {\bibinfo {author} {\bibfnamefont {M.}~\bibnamefont
  {Aguilar}} \emph {et~al.} (\bibinfo {collaboration} {AMS Collaboration}),\
  }\bibfield  {title} {\bibinfo {title} {Properties of neon, magnesium, and
  silicon primary cosmic rays results from the alpha magnetic spectrometer},\
  }\href {https://doi.org/10.1103/PhysRevLett.124.211102} {\bibfield  {journal}
  {\bibinfo  {journal} {Phys. Rev. Lett.}\ }\textbf {\bibinfo {volume} {124}},\
  \bibinfo {pages} {211102} (\bibinfo {year} {2020})}\BibitemShut {NoStop}%
\bibitem [{\citenamefont {Aguilar}\ \emph {et~al.}(2016)\citenamefont {Aguilar}
  \emph {et~al.}}]{PhysRevLett.117.231102}%
  \BibitemOpen
  \bibfield  {author} {\bibinfo {author} {\bibfnamefont {M.}~\bibnamefont
  {Aguilar}} \emph {et~al.} (\bibinfo {collaboration} {AMS Collaboration}),\
  }\bibfield  {title} {\bibinfo {title} {{Precision Measurement of the Boron to
  Carbon Flux Ratio in Cosmic Rays from 1.9 GV to 2.6 TV with the Alpha
  Magnetic Spectrometer on the International Space Station}},\ }\href
  {https://doi.org/10.1103/PhysRevLett.117.231102} {\bibfield  {journal}
  {\bibinfo  {journal} {Phys. Rev. Lett.}\ }\textbf {\bibinfo {volume} {117}},\
  \bibinfo {pages} {231102} (\bibinfo {year} {2016})}\BibitemShut {NoStop}%
\bibitem [{\citenamefont {Aguilar}\ \emph
  {et~al.}(2018{\natexlab{a}})\citenamefont {Aguilar} \emph
  {et~al.}}]{Aguilar:2018njt}%
  \BibitemOpen
  \bibfield  {author} {\bibinfo {author} {\bibfnamefont {M.}~\bibnamefont
  {Aguilar}} \emph {et~al.} (\bibinfo {collaboration} {AMS}),\ }\bibfield
  {title} {\bibinfo {title} {{Observation of New Properties of Secondary Cosmic
  Rays Lithium, Beryllium, and Boron by the Alpha Magnetic Spectrometer on the
  International Space Station}},\ }\href
  {https://doi.org/10.1103/PhysRevLett.120.021101} {\bibfield  {journal}
  {\bibinfo  {journal} {Phys. Rev. Lett.}\ }\textbf {\bibinfo {volume} {120}},\
  \bibinfo {pages} {021101} (\bibinfo {year} {2018}{\natexlab{a}})}\BibitemShut
  {NoStop}%
\bibitem [{\citenamefont {Aguilar}\ \emph
  {et~al.}(2018{\natexlab{b}})\citenamefont {Aguilar} \emph
  {et~al.}}]{2018PhRvL.121e1103A}%
  \BibitemOpen
  \bibfield  {author} {\bibinfo {author} {\bibfnamefont {M.}~\bibnamefont
  {Aguilar}} \emph {et~al.} (\bibinfo {collaboration} {AMS Collaboration}),\
  }\bibfield  {title} {\bibinfo {title} {{Precision Measurement of Cosmic-Ray
  Nitrogen and its Primary and Secondary Components with the Alpha Magnetic
  Spectrometer on the International Space Station}},\ }\href
  {https://doi.org/10.1103/PhysRevLett.121.051103} {\bibfield  {journal}
  {\bibinfo  {journal} {\prl}\ }\textbf {\bibinfo {volume} {121}},\ \bibinfo
  {eid} {051103} (\bibinfo {year} {2018}{\natexlab{b}})}\BibitemShut {NoStop}%
\bibitem [{\citenamefont {An}\ \emph {et~al.}(2019)\citenamefont {An} \emph
  {et~al.}}]{An:2019wcw}%
  \BibitemOpen
  \bibfield  {author} {\bibinfo {author} {\bibfnamefont {Q.}~\bibnamefont {An}}
  \emph {et~al.} (\bibinfo {collaboration} {DAMPE}),\ }\bibfield  {title}
  {\bibinfo {title} {{Measurement of the cosmic-ray proton spectrum from 40 GeV
  to 100 TeV with the DAMPE satellite}},\ }\href
  {https://doi.org/10.1126/sciadv.aax3793} {\bibfield  {journal} {\bibinfo
  {journal} {Sci. Adv.}\ }\textbf {\bibinfo {volume} {5}},\ \bibinfo {pages}
  {eaax3793} (\bibinfo {year} {2019})},\ \Eprint
  {https://arxiv.org/abs/1909.12860} {arXiv:1909.12860 [astro-ph.HE]}
  \BibitemShut {NoStop}%
\bibitem [{\citenamefont {Panov}\ \emph {et~al.}(2009)\citenamefont {Panov}
  \emph {et~al.}}]{Panov:2011ak}%
  \BibitemOpen
  \bibfield  {author} {\bibinfo {author} {\bibfnamefont {A.}~\bibnamefont
  {Panov}} \emph {et~al.},\ }\bibfield  {title} {\bibinfo {title} {{Energy
  Spectra of Abundant Nuclei of Primary Cosmic Rays from the Data of ATIC-2
  Experiment: Final Results}},\ }\href
  {https://doi.org/10.3103/S1062873809050098} {\bibfield  {journal} {\bibinfo
  {journal} {Bull. Russ. Acad. Sci. Phys.}\ }\textbf {\bibinfo {volume} {73}},\
  \bibinfo {pages} {564} (\bibinfo {year} {2009})},\ \Eprint
  {https://arxiv.org/abs/1101.3246} {arXiv:1101.3246 [astro-ph.HE]}
  \BibitemShut {NoStop}%
\bibitem [{\citenamefont {Atkin}\ \emph {et~al.}(2018)\citenamefont {Atkin}
  \emph {et~al.}}]{Atkin:2018wsp}%
  \BibitemOpen
  \bibfield  {author} {\bibinfo {author} {\bibfnamefont {E.}~\bibnamefont
  {Atkin}} \emph {et~al.},\ }\bibfield  {title} {\bibinfo {title} {{New
  Universal Cosmic-Ray Knee near a Magnetic Rigidity of 10 TV with the NUCLEON
  Space Observatory}},\ }\href {https://doi.org/10.1134/S0021364018130015}
  {\bibfield  {journal} {\bibinfo  {journal} {JETP Lett.}\ }\textbf {\bibinfo
  {volume} {108}},\ \bibinfo {pages} {5} (\bibinfo {year} {2018})},\ \Eprint
  {https://arxiv.org/abs/1805.07119} {arXiv:1805.07119 [astro-ph.HE]}
  \BibitemShut {NoStop}%
\bibitem [{\citenamefont {Aguilar}\ \emph
  {et~al.}(2014{\natexlab{a}})\citenamefont {Aguilar} \emph
  {et~al.}}]{2014PhRvL.113l1102A}%
  \BibitemOpen
  \bibfield  {author} {\bibinfo {author} {\bibfnamefont {M.}~\bibnamefont
  {Aguilar}} \emph {et~al.} (\bibinfo {collaboration} {AMS Collaboration}),\
  }\bibfield  {title} {\bibinfo {title} {{Electron and Positron Fluxes in
  Primary Cosmic Rays Measured with the Alpha Magnetic Spectrometer on the
  International Space Station}},\ }\href
  {https://doi.org/10.1103/PhysRevLett.113.121102} {\bibfield  {journal}
  {\bibinfo  {journal} {\prl}\ }\textbf {\bibinfo {volume} {113}},\ \bibinfo
  {eid} {121102} (\bibinfo {year} {2014}{\natexlab{a}})}\BibitemShut {NoStop}%
\bibitem [{\citenamefont {Aharonian}\ \emph {et~al.}(2009)\citenamefont
  {Aharonian} \emph {et~al.}}]{Aharonian:2009ah}%
  \BibitemOpen
  \bibfield  {author} {\bibinfo {author} {\bibfnamefont {F.}~\bibnamefont
  {Aharonian}} \emph {et~al.} (\bibinfo {collaboration} {H.E.S.S.}),\
  }\bibfield  {title} {\bibinfo {title} {{Probing the ATIC peak in the
  cosmic-ray electron spectrum with H.E.S.S}},\ }\href
  {https://doi.org/10.1051/0004-6361/200913323} {\bibfield  {journal} {\bibinfo
   {journal} {Astron. Astrophys.}\ }\textbf {\bibinfo {volume} {508}},\
  \bibinfo {pages} {561} (\bibinfo {year} {2009})},\ \Eprint
  {https://arxiv.org/abs/0905.0105} {arXiv:0905.0105 [astro-ph.HE]}
  \BibitemShut {NoStop}%
\bibitem [{\citenamefont {{Kerszberg}}(2017)}]{kerszberg_ICRC}%
  \BibitemOpen
  \bibfield  {author} {\bibinfo {author} {\bibfnamefont {D.}~\bibnamefont
  {{Kerszberg}}},\ }\bibfield  {title} {\bibinfo {title} {The cosmic-ray
  electron spectrum measured with {H.E.S.S.}},\ }\href@noop {} {\bibfield
  {journal} {\bibinfo  {journal} {International Cosmic Ray Conference, [CRI215]
  (2017)}\ } (\bibinfo {year} {2017})}\BibitemShut {NoStop}%
\bibitem [{\citenamefont {Adriani}\ \emph {et~al.}(2018)\citenamefont {Adriani}
  \emph {et~al.}}]{Adriani:2018ktz}%
  \BibitemOpen
  \bibfield  {author} {\bibinfo {author} {\bibfnamefont {O.}~\bibnamefont
  {Adriani}} \emph {et~al.},\ }\bibfield  {title} {\bibinfo {title} {{Extended
  Measurement of the Cosmic-Ray Electron and Positron Spectrum from 11 GeV to
  4.8 TeV with the Calorimetric Electron Telescope on the International Space
  Station}},\ }\href {https://doi.org/10.1103/PhysRevLett.120.261102}
  {\bibfield  {journal} {\bibinfo  {journal} {Phys. Rev. Lett.}\ }\textbf
  {\bibinfo {volume} {120}},\ \bibinfo {pages} {261102} (\bibinfo {year}
  {2018})},\ \Eprint {https://arxiv.org/abs/1806.09728} {arXiv:1806.09728
  [astro-ph.HE]} \BibitemShut {NoStop}%
\bibitem [{\citenamefont {Ambrosi}\ \emph {et~al.}(2017)\citenamefont {Ambrosi}
  \emph {et~al.}}]{Ambrosi:2017wek}%
  \BibitemOpen
  \bibfield  {author} {\bibinfo {author} {\bibfnamefont {G.}~\bibnamefont
  {Ambrosi}} \emph {et~al.} (\bibinfo {collaboration} {DAMPE}),\ }\bibfield
  {title} {\bibinfo {title} {{Direct detection of a break in the
  teraelectronvolt cosmic-ray spectrum of electrons and positrons}},\ }\href
  {https://doi.org/10.1038/nature24475} {\bibfield  {journal} {\bibinfo
  {journal} {Nature}\ }\textbf {\bibinfo {volume} {552}},\ \bibinfo {pages}
  {63} (\bibinfo {year} {2017})},\ \Eprint {https://arxiv.org/abs/1711.10981}
  {arXiv:1711.10981 [astro-ph.HE]} \BibitemShut {NoStop}%
\bibitem [{\citenamefont {Recchia}\ \emph {et~al.}(2019)\citenamefont
  {Recchia}, \citenamefont {Gabici}, \citenamefont {Aharonian},\ and\
  \citenamefont {Vink}}]{Recchia:2018jun}%
  \BibitemOpen
  \bibfield  {author} {\bibinfo {author} {\bibfnamefont {S.}~\bibnamefont
  {Recchia}}, \bibinfo {author} {\bibfnamefont {S.}~\bibnamefont {Gabici}},
  \bibinfo {author} {\bibfnamefont {F.~A.}\ \bibnamefont {Aharonian}},\ and\
  \bibinfo {author} {\bibfnamefont {J.}~\bibnamefont {Vink}},\ }\bibfield
  {title} {\bibinfo {title} {{Local fading accelerator and the origin of TeV
  cosmic ray electrons}},\ }\href {https://doi.org/10.1103/PhysRevD.99.103022}
  {\bibfield  {journal} {\bibinfo  {journal} {Phys. Rev.}\ }\textbf {\bibinfo
  {volume} {D99}},\ \bibinfo {pages} {103022} (\bibinfo {year} {2019})},\
  \Eprint {https://arxiv.org/abs/1811.07551} {arXiv:1811.07551 [astro-ph.HE]}
  \BibitemShut {NoStop}%
%%CITATION = ARXIV:1811.07551;%%
\bibitem [{\citenamefont {Fornieri}\ \emph {et~al.}(2020)\citenamefont
  {Fornieri}, \citenamefont {Gaggero},\ and\ \citenamefont
  {Grasso}}]{Fornieri:2019ddi}%
  \BibitemOpen
  \bibfield  {author} {\bibinfo {author} {\bibfnamefont {O.}~\bibnamefont
  {Fornieri}}, \bibinfo {author} {\bibfnamefont {D.}~\bibnamefont {Gaggero}},\
  and\ \bibinfo {author} {\bibfnamefont {D.}~\bibnamefont {Grasso}},\
  }\bibfield  {title} {\bibinfo {title} {{Features in cosmic-ray lepton data
  unveil the properties of nearby cosmic accelerators}},\ }\href
  {https://doi.org/10.1088/1475-7516/2020/02/009} {\bibfield  {journal}
  {\bibinfo  {journal} {JCAP}\ }\textbf {\bibinfo {volume} {02}},\ \bibinfo
  {pages} {009}},\ \Eprint {https://arxiv.org/abs/1907.03696} {arXiv:1907.03696
  [astro-ph.HE]} \BibitemShut {NoStop}%
\bibitem [{\citenamefont {Manconi}\ \emph {et~al.}(2019)\citenamefont
  {Manconi}, \citenamefont {Di~Mauro},\ and\ \citenamefont
  {Donato}}]{Manconi:2018azw}%
  \BibitemOpen
  \bibfield  {author} {\bibinfo {author} {\bibfnamefont {S.}~\bibnamefont
  {Manconi}}, \bibinfo {author} {\bibfnamefont {M.}~\bibnamefont {Di~Mauro}},\
  and\ \bibinfo {author} {\bibfnamefont {F.}~\bibnamefont {Donato}},\
  }\bibfield  {title} {\bibinfo {title} {{Multi-messenger constraints to the
  local emission of cosmic-ray electrons}},\ }\href
  {https://doi.org/10.1088/1475-7516/2019/04/024} {\bibfield  {journal}
  {\bibinfo  {journal} {JCAP}\ }\textbf {\bibinfo {volume} {04}},\ \bibinfo
  {pages} {024}},\ \Eprint {https://arxiv.org/abs/1803.01009} {arXiv:1803.01009
  [astro-ph.HE]} \BibitemShut {NoStop}%
\bibitem [{\citenamefont {{Vladimirov}}\ \emph {et~al.}(2012)\citenamefont
  {{Vladimirov}}, \citenamefont {{J{\'o}hannesson}}, \citenamefont
  {{Moskalenko}},\ and\ \citenamefont {{Porter}}}]{2012ApJ...752...68V}%
  \BibitemOpen
  \bibfield  {author} {\bibinfo {author} {\bibfnamefont {A.~E.}\ \bibnamefont
  {{Vladimirov}}}, \bibinfo {author} {\bibfnamefont {G.}~\bibnamefont
  {{J{\'o}hannesson}}}, \bibinfo {author} {\bibfnamefont {I.~V.}\ \bibnamefont
  {{Moskalenko}}},\ and\ \bibinfo {author} {\bibfnamefont {T.~A.}\ \bibnamefont
  {{Porter}}},\ }\bibfield  {title} {\bibinfo {title} {{Testing the Origin of
  High-energy Cosmic Rays}},\ }\href
  {https://doi.org/10.1088/0004-637X/752/1/68} {\bibfield  {journal} {\bibinfo
  {journal} {\apj}\ }\textbf {\bibinfo {volume} {752}},\ \bibinfo {eid} {68}
  (\bibinfo {year} {2012})},\ \Eprint {https://arxiv.org/abs/1108.1023}
  {arXiv:1108.1023 [astro-ph.HE]} \BibitemShut {NoStop}%
\bibitem [{\citenamefont {G\'enolini}\ \emph {et~al.}(2017)\citenamefont
  {G\'enolini} \emph {et~al.}}]{Genolini:2017dfb}%
  \BibitemOpen
  \bibfield  {author} {\bibinfo {author} {\bibfnamefont {Y.}~\bibnamefont
  {G\'enolini}} \emph {et~al.},\ }\bibfield  {title} {\bibinfo {title}
  {{Indications for a high-rigidity break in the cosmic-ray diffusion
  coefficient}},\ }\href {https://doi.org/10.1103/PhysRevLett.119.241101}
  {\bibfield  {journal} {\bibinfo  {journal} {Phys. Rev. Lett.}\ }\textbf
  {\bibinfo {volume} {119}},\ \bibinfo {pages} {241101} (\bibinfo {year}
  {2017})},\ \Eprint {https://arxiv.org/abs/1706.09812} {arXiv:1706.09812
  [astro-ph.HE]} \BibitemShut {NoStop}%
\bibitem [{\citenamefont {Fang}\ \emph {et~al.}(2020)\citenamefont {Fang},
  \citenamefont {Bi},\ and\ \citenamefont {Yin}}]{Fang:2020cru}%
  \BibitemOpen
  \bibfield  {author} {\bibinfo {author} {\bibfnamefont {K.}~\bibnamefont
  {Fang}}, \bibinfo {author} {\bibfnamefont {X.-J.}\ \bibnamefont {Bi}},\ and\
  \bibinfo {author} {\bibnamefont {Yin}},\ }\bibfield  {title} {\bibinfo
  {title} {{DAMPE proton spectrum indicates a slow-diffusion zone in the nearby
  ISM}},\ }\href@noop {} {\  (\bibinfo {year} {2020})},\ \Eprint
  {https://arxiv.org/abs/2003.13635} {arXiv:2003.13635 [astro-ph.HE]}
  \BibitemShut {NoStop}%
\bibitem [{\citenamefont {{Brahimi}}\ \emph {et~al.}(2020)\citenamefont
  {{Brahimi}}, \citenamefont {{Marcowith}},\ and\ \citenamefont
  {{Ptuskin}}}]{2020A&A...633A..72B}%
  \BibitemOpen
  \bibfield  {author} {\bibinfo {author} {\bibfnamefont {L.}~\bibnamefont
  {{Brahimi}}}, \bibinfo {author} {\bibfnamefont {A.}~\bibnamefont
  {{Marcowith}}},\ and\ \bibinfo {author} {\bibfnamefont {V.~S.}\ \bibnamefont
  {{Ptuskin}}},\ }\bibfield  {title} {\bibinfo {title} {{Nonlinear diffusion of
  cosmic rays escaping from supernova remnants: Cold partially neutral atomic
  and molecular phases}},\ }\href {https://doi.org/10.1051/0004-6361/201936166}
  {\bibfield  {journal} {\bibinfo  {journal} {\aap}\ }\textbf {\bibinfo
  {volume} {633}},\ \bibinfo {eid} {A72} (\bibinfo {year} {2020})},\ \Eprint
  {https://arxiv.org/abs/1909.04530} {arXiv:1909.04530 [astro-ph.HE]}
  \BibitemShut {NoStop}%
\bibitem [{\citenamefont {{Nava}}\ \emph {et~al.}(2016)\citenamefont {{Nava}},
  \citenamefont {{Gabici}}, \citenamefont {{Marcowith}}, \citenamefont
  {{Morlino}},\ and\ \citenamefont {{Ptuskin}}}]{2016MNRAS.461.3552N}%
  \BibitemOpen
  \bibfield  {author} {\bibinfo {author} {\bibfnamefont {L.}~\bibnamefont
  {{Nava}}}, \bibinfo {author} {\bibfnamefont {S.}~\bibnamefont {{Gabici}}},
  \bibinfo {author} {\bibfnamefont {A.}~\bibnamefont {{Marcowith}}}, \bibinfo
  {author} {\bibfnamefont {G.}~\bibnamefont {{Morlino}}},\ and\ \bibinfo
  {author} {\bibfnamefont {V.~S.}\ \bibnamefont {{Ptuskin}}},\ }\bibfield
  {title} {\bibinfo {title} {{Non-linear diffusion of cosmic rays escaping from
  supernova remnants - I. The effect of neutrals}},\ }\href
  {https://doi.org/10.1093/mnras/stw1592} {\bibfield  {journal} {\bibinfo
  {journal} {\mnras}\ }\textbf {\bibinfo {volume} {461}},\ \bibinfo {pages}
  {3552} (\bibinfo {year} {2016})},\ \Eprint {https://arxiv.org/abs/1606.06902}
  {arXiv:1606.06902 [astro-ph.HE]} \BibitemShut {NoStop}%
\bibitem [{\citenamefont {Liu}\ \emph {et~al.}(2019)\citenamefont {Liu},
  \citenamefont {Guo},\ and\ \citenamefont {Yuan}}]{Liu:2018fjy}%
  \BibitemOpen
  \bibfield  {author} {\bibinfo {author} {\bibfnamefont {W.}~\bibnamefont
  {Liu}}, \bibinfo {author} {\bibfnamefont {Y.-Q.}\ \bibnamefont {Guo}},\ and\
  \bibinfo {author} {\bibfnamefont {Q.}~\bibnamefont {Yuan}},\ }\bibfield
  {title} {\bibinfo {title} {{Indication of nearby source signatures of cosmic
  rays from energy spectra and anisotropies}},\ }\href
  {https://doi.org/10.1088/1475-7516/2019/10/010} {\bibfield  {journal}
  {\bibinfo  {journal} {JCAP}\ }\textbf {\bibinfo {volume} {10}},\ \bibinfo
  {pages} {010}},\ \Eprint {https://arxiv.org/abs/1812.09673} {arXiv:1812.09673
  [astro-ph.HE]} \BibitemShut {NoStop}%
\bibitem [{\citenamefont {Yuan}\ \emph {et~al.}(2021)\citenamefont {Yuan},
  \citenamefont {Qiao}, \citenamefont {Guo}, \citenamefont {Fan},\ and\
  \citenamefont {Bi}}]{Yuan:2020ciu}%
  \BibitemOpen
  \bibfield  {author} {\bibinfo {author} {\bibfnamefont {Q.}~\bibnamefont
  {Yuan}}, \bibinfo {author} {\bibfnamefont {B.-Q.}\ \bibnamefont {Qiao}},
  \bibinfo {author} {\bibfnamefont {Y.-Q.}\ \bibnamefont {Guo}}, \bibinfo
  {author} {\bibfnamefont {Y.-Z.}\ \bibnamefont {Fan}},\ and\ \bibinfo {author}
  {\bibfnamefont {X.-J.}\ \bibnamefont {Bi}},\ }\bibfield  {title} {\bibinfo
  {title} {{Nearby source interpretation of differences among light and medium
  composition spectra in cosmic rays}},\ }\href
  {https://doi.org/10.1007/s11467-020-0990-4} {\bibfield  {journal} {\bibinfo
  {journal} {Front. Phys. (Beijing)}\ }\textbf {\bibinfo {volume} {16}},\
  \bibinfo {pages} {24501} (\bibinfo {year} {2021})},\ \Eprint
  {https://arxiv.org/abs/2007.01768} {arXiv:2007.01768 [astro-ph.HE]}
  \BibitemShut {NoStop}%
\bibitem [{\citenamefont {{Malkov}}\ and\ \citenamefont
  {{Moskalenko}}(2020)}]{2020arXiv201002826M}%
  \BibitemOpen
  \bibfield  {author} {\bibinfo {author} {\bibfnamefont {M.~A.}\ \bibnamefont
  {{Malkov}}}\ and\ \bibinfo {author} {\bibfnamefont {I.~V.}\ \bibnamefont
  {{Moskalenko}}},\ }\bibfield  {title} {\bibinfo {title} {{The TeV Cosmic Ray
  Bump: a Message from Epsilon Indi Star?}},\ }\href@noop {} {\bibfield
  {journal} {\bibinfo  {journal} {arXiv e-prints}\ ,\ \bibinfo {eid}
  {arXiv:2010.02826}} (\bibinfo {year} {2020})},\ \Eprint
  {https://arxiv.org/abs/2010.02826} {arXiv:2010.02826 [astro-ph.HE]}
  \BibitemShut {NoStop}%
\bibitem [{\citenamefont {{Yuan}}\ \emph {et~al.}(2020)\citenamefont {{Yuan}},
  \citenamefont {{Qiao}}, \citenamefont {{Guo}}, \citenamefont {{Fan}},\ and\
  \citenamefont {{Bi}}}]{2020FrPhy..1624501Y}%
  \BibitemOpen
  \bibfield  {author} {\bibinfo {author} {\bibfnamefont {Q.}~\bibnamefont
  {{Yuan}}}, \bibinfo {author} {\bibfnamefont {B.-Q.}\ \bibnamefont {{Qiao}}},
  \bibinfo {author} {\bibfnamefont {Y.-Q.}\ \bibnamefont {{Guo}}}, \bibinfo
  {author} {\bibfnamefont {Y.-Z.}\ \bibnamefont {{Fan}}},\ and\ \bibinfo
  {author} {\bibfnamefont {X.-J.}\ \bibnamefont {{Bi}}},\ }\bibfield  {title}
  {\bibinfo {title} {{Nearby source interpretation of differences among light
  and medium composition spectra in cosmic rays}},\ }\href
  {https://doi.org/10.1007/s11467-020-0990-4} {\bibfield  {journal} {\bibinfo
  {journal} {Frontiers of Physics}\ }\textbf {\bibinfo {volume} {16}},\
  \bibinfo {eid} {24501} (\bibinfo {year} {2020})},\ \Eprint
  {https://arxiv.org/abs/2007.01768} {arXiv:2007.01768 [astro-ph.HE]}
  \BibitemShut {NoStop}%
\bibitem [{\citenamefont {Fornieri}\ \emph {et~al.}()\citenamefont {Fornieri},
  \citenamefont {Gaggero}, \citenamefont {Guberman}, \citenamefont {Brahimi},
  \citenamefont {De~La Torre~Luque},\ and\ \citenamefont
  {Marcowith}}]{Fornieri:2021PRL}%
  \BibitemOpen
  \bibfield  {author} {\bibinfo {author} {\bibfnamefont {O.}~\bibnamefont
  {Fornieri}}, \bibinfo {author} {\bibfnamefont {D.}~\bibnamefont {Gaggero}},
  \bibinfo {author} {\bibfnamefont {D.}~\bibnamefont {Guberman}}, \bibinfo
  {author} {\bibfnamefont {L.}~\bibnamefont {Brahimi}}, \bibinfo {author}
  {\bibfnamefont {P.}~\bibnamefont {De~La Torre~Luque}},\ and\ \bibinfo
  {author} {\bibfnamefont {A.}~\bibnamefont {Marcowith}},\ }\bibfield  {title}
  {\bibinfo {title} {{On the imprint of a local accelerator on three different
  cosmic-ray observables}},\ }\href@noop {} {\bibinfo  {journal} {submitted to
  PRL}\ }\BibitemShut {NoStop}%
\bibitem [{\citenamefont {Evoli}\ \emph {et~al.}(2008)\citenamefont {Evoli},
  \citenamefont {Gaggero}, \citenamefont {Grasso},\ and\ \citenamefont
  {Maccione}}]{Evoli:2008dv}%
  \BibitemOpen
\bibfield  {journal} {  }\bibfield  {author} {\bibinfo {author} {\bibfnamefont
  {C.}~\bibnamefont {Evoli}}, \bibinfo {author} {\bibfnamefont
  {D.}~\bibnamefont {Gaggero}}, \bibinfo {author} {\bibfnamefont
  {D.}~\bibnamefont {Grasso}},\ and\ \bibinfo {author} {\bibfnamefont
  {L.}~\bibnamefont {Maccione}},\ }\bibfield  {title} {\bibinfo {title}
  {{Cosmic-Ray Nuclei, Antiprotons and Gamma-rays in the Galaxy: a New
  Diffusion Model}},\ }\href {https://doi.org/10.1088/1475-7516/2008/10/018}
  {\bibfield  {journal} {\bibinfo  {journal} {JCAP}\ }\textbf {\bibinfo
  {volume} {0810}},\ \bibinfo {pages} {018}},\ \Eprint
  {https://arxiv.org/abs/0807.4730} {arXiv:0807.4730 [astro-ph]} \BibitemShut
  {NoStop}%
%%CITATION = ARXIV:0807.4730;%%
\bibitem [{\citenamefont {{Evoli}}\ \emph {et~al.}(2017)\citenamefont
  {{Evoli}}, \citenamefont {{Gaggero}}, \citenamefont {{Vittino}},
  \citenamefont {{Di Bernardo}}, \citenamefont {{Di Mauro}}, \citenamefont
  {{Ligorini}}, \citenamefont {{Ullio}},\ and\ \citenamefont
  {{Grasso}}}]{2017JCAP...02..015E}%
  \BibitemOpen
  \bibfield  {author} {\bibinfo {author} {\bibfnamefont {C.}~\bibnamefont
  {{Evoli}}}, \bibinfo {author} {\bibfnamefont {D.}~\bibnamefont {{Gaggero}}},
  \bibinfo {author} {\bibfnamefont {A.}~\bibnamefont {{Vittino}}}, \bibinfo
  {author} {\bibfnamefont {G.}~\bibnamefont {{Di Bernardo}}}, \bibinfo {author}
  {\bibfnamefont {M.}~\bibnamefont {{Di Mauro}}}, \bibinfo {author}
  {\bibfnamefont {A.}~\bibnamefont {{Ligorini}}}, \bibinfo {author}
  {\bibfnamefont {P.}~\bibnamefont {{Ullio}}},\ and\ \bibinfo {author}
  {\bibfnamefont {D.}~\bibnamefont {{Grasso}}},\ }\bibfield  {title} {\bibinfo
  {title} {{Cosmic-ray propagation with DRAGON2: I. numerical solver and
  astrophysical ingredients}},\ }\href
  {https://doi.org/10.1088/1475-7516/2017/02/015} {\bibfield  {journal}
  {\bibinfo  {journal} {\jcap}\ }\textbf {\bibinfo {volume} {2}},\ \bibinfo
  {eid} {015} (\bibinfo {year} {2017})},\ \Eprint
  {https://arxiv.org/abs/1607.07886} {arXiv:1607.07886 [astro-ph.HE]}
  \BibitemShut {NoStop}%
\bibitem [{\citenamefont {Tomassetti}(2012)}]{Tomassetti:2012ga}%
  \BibitemOpen
  \bibfield  {author} {\bibinfo {author} {\bibfnamefont {N.}~\bibnamefont
  {Tomassetti}},\ }\bibfield  {title} {\bibinfo {title} {{Origin of the
  Cosmic-Ray Spectral Hardening}},\ }\href
  {https://doi.org/10.1088/2041-8205/752/1/L13} {\bibfield  {journal} {\bibinfo
   {journal} {Astrophys. J. Lett.}\ }\textbf {\bibinfo {volume} {752}},\
  \bibinfo {pages} {L13} (\bibinfo {year} {2012})},\ \Eprint
  {https://arxiv.org/abs/1204.4492} {arXiv:1204.4492 [astro-ph.HE]}
  \BibitemShut {NoStop}%
\bibitem [{\citenamefont {Blasi}\ \emph {et~al.}(2012)\citenamefont {Blasi},
  \citenamefont {Amato},\ and\ \citenamefont {Serpico}}]{Blasi:2012yr}%
  \BibitemOpen
  \bibfield  {author} {\bibinfo {author} {\bibfnamefont {P.}~\bibnamefont
  {Blasi}}, \bibinfo {author} {\bibfnamefont {E.}~\bibnamefont {Amato}},\ and\
  \bibinfo {author} {\bibfnamefont {P.~D.}\ \bibnamefont {Serpico}},\
  }\bibfield  {title} {\bibinfo {title} {{Spectral breaks as a signature of
  cosmic ray induced turbulence in the Galaxy}},\ }\href
  {https://doi.org/10.1103/PhysRevLett.109.061101} {\bibfield  {journal}
  {\bibinfo  {journal} {Phys. Rev. Lett.}\ }\textbf {\bibinfo {volume} {109}},\
  \bibinfo {pages} {061101} (\bibinfo {year} {2012})},\ \Eprint
  {https://arxiv.org/abs/1207.3706} {arXiv:1207.3706 [astro-ph.HE]}
  \BibitemShut {NoStop}%
\bibitem [{\citenamefont {Yan}\ and\ \citenamefont
  {Lazarian}(2002)}]{Yan:2002qm}%
  \BibitemOpen
  \bibfield  {author} {\bibinfo {author} {\bibfnamefont {H.}~\bibnamefont
  {Yan}}\ and\ \bibinfo {author} {\bibfnamefont {A.}~\bibnamefont {Lazarian}},\
  }\bibfield  {title} {\bibinfo {title} {{Scattering of cosmic rays by
  magnetohydrodynamic interstellar turbulence}},\ }\href
  {https://doi.org/10.1103/PhysRevLett.89.281102} {\bibfield  {journal}
  {\bibinfo  {journal} {Phys. Rev. Lett.}\ }\textbf {\bibinfo {volume} {89}},\
  \bibinfo {pages} {281102} (\bibinfo {year} {2002})},\ \Eprint
  {https://arxiv.org/abs/astro-ph/0205285} {arXiv:astro-ph/0205285 [astro-ph]}
  \BibitemShut {NoStop}%
%%CITATION = ASTRO-PH/0205285;%%
\bibitem [{\citenamefont {Yan}\ and\ \citenamefont
  {Lazarian}(2004)}]{Yan:2004aq}%
  \BibitemOpen
  \bibfield  {author} {\bibinfo {author} {\bibfnamefont {H.}~\bibnamefont
  {Yan}}\ and\ \bibinfo {author} {\bibfnamefont {A.}~\bibnamefont {Lazarian}},\
  }\bibfield  {title} {\bibinfo {title} {{Cosmic ray scattering and streaming
  in compressible magnetohydrodynamic turbulence}},\ }\href
  {https://doi.org/10.1086/423733} {\bibfield  {journal} {\bibinfo  {journal}
  {Astrophys. J.}\ }\textbf {\bibinfo {volume} {614}},\ \bibinfo {pages} {757}
  (\bibinfo {year} {2004})},\ \Eprint {https://arxiv.org/abs/astro-ph/0408172}
  {arXiv:astro-ph/0408172 [astro-ph]} \BibitemShut {NoStop}%
%%CITATION = ASTRO-PH/0408172;%%
\bibitem [{\citenamefont {Yan}\ and\ \citenamefont
  {Lazarian}(2008)}]{Yan:2007uc}%
  \BibitemOpen
  \bibfield  {author} {\bibinfo {author} {\bibfnamefont {H.}~\bibnamefont
  {Yan}}\ and\ \bibinfo {author} {\bibfnamefont {A.}~\bibnamefont {Lazarian}},\
  }\bibfield  {title} {\bibinfo {title} {{Cosmic Ray Propagation: Nonlinear
  Diffusion Parallel and Perpendicular to Mean Magnetic Field}},\ }\href
  {https://doi.org/10.1086/524771} {\bibfield  {journal} {\bibinfo  {journal}
  {Astrophys. J.}\ }\textbf {\bibinfo {volume} {673}},\ \bibinfo {pages} {942}
  (\bibinfo {year} {2008})},\ \Eprint {https://arxiv.org/abs/0710.2617}
  {arXiv:0710.2617 [astro-ph]} \BibitemShut {NoStop}%
%%CITATION = ARXIV:0710.2617;%%
\bibitem [{\citenamefont {Evoli}\ and\ \citenamefont
  {Yan}(2014)}]{Evoli:2013lma}%
  \BibitemOpen
  \bibfield  {author} {\bibinfo {author} {\bibfnamefont {C.}~\bibnamefont
  {Evoli}}\ and\ \bibinfo {author} {\bibfnamefont {H.}~\bibnamefont {Yan}},\
  }\bibfield  {title} {\bibinfo {title} {{Cosmic ray propagation in galactic
  turbulence}},\ }\href {https://doi.org/10.1088/0004-637X/782/1/36} {\bibfield
   {journal} {\bibinfo  {journal} {Astrophys. J.}\ }\textbf {\bibinfo {volume}
  {782}},\ \bibinfo {pages} {36} (\bibinfo {year} {2014})},\ \Eprint
  {https://arxiv.org/abs/1310.5732} {arXiv:1310.5732 [astro-ph.HE]}
  \BibitemShut {NoStop}%
\bibitem [{\citenamefont {Fornieri}\ \emph {et~al.}(2021)\citenamefont
  {Fornieri}, \citenamefont {Gaggero}, \citenamefont {Cerri}, \citenamefont
  {Luque},\ and\ \citenamefont {Gabici}}]{10.1093/mnras/stab355}%
  \BibitemOpen
  \bibfield  {author} {\bibinfo {author} {\bibfnamefont {O.}~\bibnamefont
  {Fornieri}}, \bibinfo {author} {\bibfnamefont {D.}~\bibnamefont {Gaggero}},
  \bibinfo {author} {\bibfnamefont {S.~S.}\ \bibnamefont {Cerri}}, \bibinfo
  {author} {\bibfnamefont {P.~D. L.~T.}\ \bibnamefont {Luque}},\ and\ \bibinfo
  {author} {\bibfnamefont {S.}~\bibnamefont {Gabici}},\ }\bibfield  {title}
  {\bibinfo {title} {{The theory of cosmic-ray scattering on pre-existing MHD
  modes meets data}},\ }\bibfield  {journal} {\bibinfo  {journal} {Monthly
  Notices of the Royal Astronomical Society}\ }\href
  {https://doi.org/10.1093/mnras/stab355} {10.1093/mnras/stab355} (\bibinfo
  {year} {2021}),\ \bibinfo {note} {stab355},\ \Eprint
  {https://arxiv.org/abs/https://academic.oup.com/mnras/advance-article-pdf/doi/10.1093/mnras/stab355/36219839/stab355.pdf}
  {https://academic.oup.com/mnras/advance-article-pdf/doi/10.1093/mnras/stab355/36219839/stab355.pdf}
  \BibitemShut {NoStop}%
\bibitem [{\citenamefont {Farmer}\ and\ \citenamefont
  {Goldreich}(2004)}]{Farmer:2003mz}%
  \BibitemOpen
  \bibfield  {author} {\bibinfo {author} {\bibfnamefont {A.~J.}\ \bibnamefont
  {Farmer}}\ and\ \bibinfo {author} {\bibfnamefont {P.}~\bibnamefont
  {Goldreich}},\ }\bibfield  {title} {\bibinfo {title} {{Wave damping by MHD
  turbulence and its effect upon cosmic ray propagation in the ISM}},\ }\href
  {https://doi.org/10.1086/382040} {\bibfield  {journal} {\bibinfo  {journal}
  {Astrophys. J.}\ }\textbf {\bibinfo {volume} {604}},\ \bibinfo {pages} {671}
  (\bibinfo {year} {2004})},\ \Eprint {https://arxiv.org/abs/astro-ph/0311400}
  {arXiv:astro-ph/0311400} \BibitemShut {NoStop}%
\bibitem [{\citenamefont {Feng}\ \emph {et~al.}(2016)\citenamefont {Feng},
  \citenamefont {Tomassetti},\ and\ \citenamefont {Oliva}}]{Feng:2016loc}%
  \BibitemOpen
  \bibfield  {author} {\bibinfo {author} {\bibfnamefont {J.}~\bibnamefont
  {Feng}}, \bibinfo {author} {\bibfnamefont {N.}~\bibnamefont {Tomassetti}},\
  and\ \bibinfo {author} {\bibfnamefont {A.}~\bibnamefont {Oliva}},\ }\bibfield
   {title} {\bibinfo {title} {{Bayesian analysis of spatial-dependent
  cosmic-ray propagation: astrophysical background of antiprotons and
  positrons}},\ }\href {https://doi.org/10.1103/PhysRevD.94.123007} {\bibfield
  {journal} {\bibinfo  {journal} {Phys. Rev. D}\ }\textbf {\bibinfo {volume}
  {94}},\ \bibinfo {pages} {123007} (\bibinfo {year} {2016})},\ \Eprint
  {https://arxiv.org/abs/1610.06182} {arXiv:1610.06182 [astro-ph.HE]}
  \BibitemShut {NoStop}%
\bibitem [{\citenamefont {{G{\'e}nolini}}\ \emph {et~al.}(2019)\citenamefont
  {{G{\'e}nolini}}, \citenamefont {{Boudaud}}, \citenamefont {{Batista}},
  \citenamefont {{Caroff}}, \citenamefont {{Derome}}, \citenamefont
  {{Lavalle}}, \citenamefont {{Marcowith}}, \citenamefont {{Maurin}},
  \citenamefont {{Poireau}}, \citenamefont {{Poulin}}, \citenamefont
  {{Rosier}}, \citenamefont {{Salati}}, \citenamefont {{Serpico}},\ and\
  \citenamefont {{Vecchi}}}]{2019PhRvD..99l3028G}%
  \BibitemOpen
  \bibfield  {author} {\bibinfo {author} {\bibfnamefont {Y.}~\bibnamefont
  {{G{\'e}nolini}}}, \bibinfo {author} {\bibfnamefont {M.}~\bibnamefont
  {{Boudaud}}}, \bibinfo {author} {\bibfnamefont {P.~I.}\ \bibnamefont
  {{Batista}}}, \bibinfo {author} {\bibfnamefont {S.}~\bibnamefont {{Caroff}}},
  \bibinfo {author} {\bibfnamefont {L.}~\bibnamefont {{Derome}}}, \bibinfo
  {author} {\bibfnamefont {J.}~\bibnamefont {{Lavalle}}}, \bibinfo {author}
  {\bibfnamefont {A.}~\bibnamefont {{Marcowith}}}, \bibinfo {author}
  {\bibfnamefont {D.}~\bibnamefont {{Maurin}}}, \bibinfo {author}
  {\bibfnamefont {V.}~\bibnamefont {{Poireau}}}, \bibinfo {author}
  {\bibfnamefont {V.}~\bibnamefont {{Poulin}}}, \bibinfo {author}
  {\bibfnamefont {S.}~\bibnamefont {{Rosier}}}, \bibinfo {author}
  {\bibfnamefont {P.}~\bibnamefont {{Salati}}}, \bibinfo {author}
  {\bibfnamefont {P.~D.}\ \bibnamefont {{Serpico}}},\ and\ \bibinfo {author}
  {\bibfnamefont {M.}~\bibnamefont {{Vecchi}}},\ }\bibfield  {title} {\bibinfo
  {title} {{Cosmic-ray transport from AMS-02 boron to carbon ratio data:
  Benchmark models and interpretation}},\ }\href
  {https://doi.org/10.1103/PhysRevD.99.123028} {\bibfield  {journal} {\bibinfo
  {journal} {\prd}\ }\textbf {\bibinfo {volume} {99}},\ \bibinfo {eid} {123028}
  (\bibinfo {year} {2019})},\ \Eprint {https://arxiv.org/abs/1904.08917}
  {arXiv:1904.08917 [astro-ph.HE]} \BibitemShut {NoStop}%
\bibitem [{\citenamefont {Ptuskin}\ \emph {et~al.}(2006)\citenamefont
  {Ptuskin}, \citenamefont {Moskalenko}, \citenamefont {Jones}, \citenamefont
  {Strong},\ and\ \citenamefont {Zirakashvili}}]{Ptuskin:2005ax}%
  \BibitemOpen
  \bibfield  {author} {\bibinfo {author} {\bibfnamefont {V.}~\bibnamefont
  {Ptuskin}}, \bibinfo {author} {\bibfnamefont {I.~V.}\ \bibnamefont
  {Moskalenko}}, \bibinfo {author} {\bibfnamefont {F.}~\bibnamefont {Jones}},
  \bibinfo {author} {\bibfnamefont {A.}~\bibnamefont {Strong}},\ and\ \bibinfo
  {author} {\bibfnamefont {V.}~\bibnamefont {Zirakashvili}},\ }\bibfield
  {title} {\bibinfo {title} {{Dissipation of magnetohydrodynamic waves on
  energetic particles: impact on interstellar turbulence and cosmic ray
  transport}},\ }\href {https://doi.org/10.1086/501117} {\bibfield  {journal}
  {\bibinfo  {journal} {Astrophys. J.}\ }\textbf {\bibinfo {volume} {642}},\
  \bibinfo {pages} {902} (\bibinfo {year} {2006})},\ \Eprint
  {https://arxiv.org/abs/astro-ph/0510335} {arXiv:astro-ph/0510335}
  \BibitemShut {NoStop}%
\bibitem [{\citenamefont {Reichherzer}\ \emph {et~al.}(2020)\citenamefont
  {Reichherzer}, \citenamefont {Becker~Tjus}, \citenamefont {Zweibel},
  \citenamefont {Merten},\ and\ \citenamefont
  {Pueschel}}]{Reichherzer:2019dmb}%
  \BibitemOpen
  \bibfield  {author} {\bibinfo {author} {\bibfnamefont {P.}~\bibnamefont
  {Reichherzer}}, \bibinfo {author} {\bibfnamefont {J.}~\bibnamefont
  {Becker~Tjus}}, \bibinfo {author} {\bibfnamefont {E.}~\bibnamefont
  {Zweibel}}, \bibinfo {author} {\bibfnamefont {L.}~\bibnamefont {Merten}},\
  and\ \bibinfo {author} {\bibfnamefont {M.}~\bibnamefont {Pueschel}},\
  }\bibfield  {title} {\bibinfo {title} {{Turbulence-Level Dependence of
  Cosmic-Ray Parallel Diffusion}},\ }\href
  {https://doi.org/10.1093/mnras/staa2533} {\bibfield  {journal} {\bibinfo
  {journal} {Mon. Not. Roy. Astron. Soc.}\ }\textbf {\bibinfo {volume} {498}},\
  \bibinfo {pages} {5051} (\bibinfo {year} {2020})},\ \Eprint
  {https://arxiv.org/abs/1910.07528} {arXiv:1910.07528 [astro-ph.HE]}
  \BibitemShut {NoStop}%
\bibitem [{\citenamefont {Adriani}\ \emph {et~al.}(2014)\citenamefont {Adriani}
  \emph {et~al.}}]{Adriani:2014xoa}%
  \BibitemOpen
  \bibfield  {author} {\bibinfo {author} {\bibfnamefont {O.}~\bibnamefont
  {Adriani}} \emph {et~al.},\ }\bibfield  {title} {\bibinfo {title}
  {{Measurement of boron and carbon fluxes in cosmic rays with the PAMELA
  experiment}},\ }\href {https://doi.org/10.1088/0004-637X/791/2/93} {\bibfield
   {journal} {\bibinfo  {journal} {Astrophys. J.}\ }\textbf {\bibinfo {volume}
  {791}},\ \bibinfo {pages} {93} (\bibinfo {year} {2014})},\ \Eprint
  {https://arxiv.org/abs/1407.1657} {arXiv:1407.1657 [astro-ph.HE]}
  \BibitemShut {NoStop}%
\bibitem [{\citenamefont {Cummings}\ \emph {et~al.}(2016)\citenamefont
  {Cummings}, \citenamefont {Stone}, \citenamefont {Heikkila}, \citenamefont
  {Lal}, \citenamefont {Webber}, \citenamefont {J{\'{o}}hannesson},
  \citenamefont {Moskalenko}, \citenamefont {Orlando},\ and\ \citenamefont
  {Porter}}]{Cummings_2016}%
  \BibitemOpen
  \bibfield  {author} {\bibinfo {author} {\bibfnamefont {A.~C.}\ \bibnamefont
  {Cummings}}, \bibinfo {author} {\bibfnamefont {E.~C.}\ \bibnamefont {Stone}},
  \bibinfo {author} {\bibfnamefont {B.~C.}\ \bibnamefont {Heikkila}}, \bibinfo
  {author} {\bibfnamefont {N.}~\bibnamefont {Lal}}, \bibinfo {author}
  {\bibfnamefont {W.~R.}\ \bibnamefont {Webber}}, \bibinfo {author}
  {\bibfnamefont {G.}~\bibnamefont {J{\'{o}}hannesson}}, \bibinfo {author}
  {\bibfnamefont {I.~V.}\ \bibnamefont {Moskalenko}}, \bibinfo {author}
  {\bibfnamefont {E.}~\bibnamefont {Orlando}},\ and\ \bibinfo {author}
  {\bibfnamefont {T.~A.}\ \bibnamefont {Porter}},\ }\bibfield  {title}
  {\bibinfo {title} {Galactic cosmic rays in the local interstellar medium:
  Voyager-1 observations and model results},\ }\href
  {https://doi.org/10.3847/0004-637x/831/1/18} {\bibfield  {journal} {\bibinfo
  {journal} {The Astrophysical Journal}\ }\textbf {\bibinfo {volume} {831}},\
  \bibinfo {pages} {18} (\bibinfo {year} {2016})}\BibitemShut {NoStop}%
\bibitem [{\citenamefont {{Gleeson}}\ and\ \citenamefont
  {{Axford}}(1968)}]{1968ApJ...154.1011G}%
  \BibitemOpen
  \bibfield  {author} {\bibinfo {author} {\bibfnamefont {L.~J.}\ \bibnamefont
  {{Gleeson}}}\ and\ \bibinfo {author} {\bibfnamefont {W.~I.}\ \bibnamefont
  {{Axford}}},\ }\bibfield  {title} {\bibinfo {title} {{Solar Modulation of
  Galactic Cosmic Rays}},\ }\href {https://doi.org/10.1086/149822} {\bibfield
  {journal} {\bibinfo  {journal} {\apj}\ }\textbf {\bibinfo {volume} {154}},\
  \bibinfo {pages} {1011} (\bibinfo {year} {1968})}\BibitemShut {NoStop}%
\bibitem [{\citenamefont {{Usoskin}}\ \emph {et~al.}(2005)\citenamefont
  {{Usoskin}}, \citenamefont {{Alanko-Huotari}}, \citenamefont {{Kovaltsov}},\
  and\ \citenamefont {{Mursula}}}]{2005JGRA..11012108U}%
  \BibitemOpen
  \bibfield  {author} {\bibinfo {author} {\bibfnamefont {I.~G.}\ \bibnamefont
  {{Usoskin}}}, \bibinfo {author} {\bibfnamefont {K.}~\bibnamefont
  {{Alanko-Huotari}}}, \bibinfo {author} {\bibfnamefont {G.~A.}\ \bibnamefont
  {{Kovaltsov}}},\ and\ \bibinfo {author} {\bibfnamefont {K.}~\bibnamefont
  {{Mursula}}},\ }\bibfield  {title} {\bibinfo {title} {{Heliospheric
  modulation of cosmic rays: Monthly reconstruction for 1951-2004}},\ }\href
  {https://doi.org/10.1029/2005JA011250} {\bibfield  {journal} {\bibinfo
  {journal} {Journal of Geophysical Research (Space Physics)}\ }\textbf
  {\bibinfo {volume} {110}},\ \bibinfo {eid} {A12108} (\bibinfo {year}
  {2005})}\BibitemShut {NoStop}%
\bibitem [{\citenamefont {{Usoskin}}\ \emph {et~al.}(2011)\citenamefont
  {{Usoskin}}, \citenamefont {{Bazilevskaya}},\ and\ \citenamefont
  {{Kovaltsov}}}]{2011JGRA..116.2104U}%
  \BibitemOpen
  \bibfield  {author} {\bibinfo {author} {\bibfnamefont {I.~G.}\ \bibnamefont
  {{Usoskin}}}, \bibinfo {author} {\bibfnamefont {G.~A.}\ \bibnamefont
  {{Bazilevskaya}}},\ and\ \bibinfo {author} {\bibfnamefont {G.~A.}\
  \bibnamefont {{Kovaltsov}}},\ }\bibfield  {title} {\bibinfo {title} {{Solar
  modulation parameter for cosmic rays since 1936 reconstructed from
  ground-based neutron monitors and ionization chambers}},\ }\href
  {https://doi.org/10.1029/2010JA016105} {\bibfield  {journal} {\bibinfo
  {journal} {Journal of Geophysical Research (Space Physics)}\ }\textbf
  {\bibinfo {volume} {116}},\ \bibinfo {eid} {A02104} (\bibinfo {year}
  {2011})}\BibitemShut {NoStop}%
\bibitem [{\citenamefont {Ferriere}(2001)}]{Ferriere:2001rg}%
  \BibitemOpen
  \bibfield  {author} {\bibinfo {author} {\bibfnamefont {K.~M.}\ \bibnamefont
  {Ferriere}},\ }\bibfield  {title} {\bibinfo {title} {{The interstellar
  environment of our galaxy}},\ }\href
  {https://doi.org/10.1103/RevModPhys.73.1031} {\bibfield  {journal} {\bibinfo
  {journal} {Rev. Mod. Phys.}\ }\textbf {\bibinfo {volume} {73}},\ \bibinfo
  {pages} {1031} (\bibinfo {year} {2001})},\ \Eprint
  {https://arxiv.org/abs/astro-ph/0106359} {arXiv:astro-ph/0106359 [astro-ph]}
  \BibitemShut {NoStop}%
%%CITATION = ASTRO-PH/0106359;%%
\bibitem [{\citenamefont {Ferrand}\ and\ \citenamefont
  {Safi-Harb}(2012)}]{Ferrand:2012jh}%
  \BibitemOpen
  \bibfield  {author} {\bibinfo {author} {\bibfnamefont {G.}~\bibnamefont
  {Ferrand}}\ and\ \bibinfo {author} {\bibfnamefont {S.}~\bibnamefont
  {Safi-Harb}},\ }\bibfield  {title} {\bibinfo {title} {{A Census of
  High-Energy Observations of Galactic Supernova Remnants}},\ }\href
  {https://doi.org/10.1016/j.asr.2012.02.004} {\bibfield  {journal} {\bibinfo
  {journal} {Adv. Space Res.}\ }\textbf {\bibinfo {volume} {49}},\ \bibinfo
  {pages} {1313} (\bibinfo {year} {2012})},\ \Eprint
  {https://arxiv.org/abs/1202.0245} {arXiv:1202.0245 [astro-ph.HE]}
  \BibitemShut {NoStop}%
\bibitem [{\citenamefont {{Atoyan}}\ \emph {et~al.}(1995)\citenamefont
  {{Atoyan}}, \citenamefont {{Aharonian}},\ and\ \citenamefont
  {{V{\"o}lk}}}]{1995PhRvD..52.3265A}%
  \BibitemOpen
  \bibfield  {author} {\bibinfo {author} {\bibfnamefont {A.~M.}\ \bibnamefont
  {{Atoyan}}}, \bibinfo {author} {\bibfnamefont {F.~A.}\ \bibnamefont
  {{Aharonian}}},\ and\ \bibinfo {author} {\bibfnamefont {H.~J.}\ \bibnamefont
  {{V{\"o}lk}}},\ }\bibfield  {title} {\bibinfo {title} {{Electrons and
  positrons in the galactic cosmic rays}},\ }\href
  {https://doi.org/10.1103/PhysRevD.52.3265} {\bibfield  {journal} {\bibinfo
  {journal} {\prd}\ }\textbf {\bibinfo {volume} {52}},\ \bibinfo {pages} {3265}
  (\bibinfo {year} {1995})}\BibitemShut {NoStop}%
\bibitem [{\citenamefont {{Hooper}}\ \emph {et~al.}(2017)\citenamefont
  {{Hooper}}, \citenamefont {{Cholis}}, \citenamefont {{Linden}},\ and\
  \citenamefont {{Fang}}}]{2017PhRvD..96j3013H}%
  \BibitemOpen
  \bibfield  {author} {\bibinfo {author} {\bibfnamefont {D.}~\bibnamefont
  {{Hooper}}}, \bibinfo {author} {\bibfnamefont {I.}~\bibnamefont {{Cholis}}},
  \bibinfo {author} {\bibfnamefont {T.}~\bibnamefont {{Linden}}},\ and\
  \bibinfo {author} {\bibfnamefont {K.}~\bibnamefont {{Fang}}},\ }\bibfield
  {title} {\bibinfo {title} {{HAWC observations strongly favor pulsar
  interpretations of the cosmic-ray positron excess}},\ }\href
  {https://doi.org/10.1103/PhysRevD.96.103013} {\bibfield  {journal} {\bibinfo
  {journal} {\prd}\ }\textbf {\bibinfo {volume} {96}},\ \bibinfo {eid} {103013}
  (\bibinfo {year} {2017})},\ \Eprint {https://arxiv.org/abs/1702.08436}
  {arXiv:1702.08436 [astro-ph.HE]} \BibitemShut {NoStop}%
\bibitem [{\citenamefont {{Evoli}}\ \emph
  {et~al.}(2020{\natexlab{a}})\citenamefont {{Evoli}}, \citenamefont {{Blasi}},
  \citenamefont {{Amato}},\ and\ \citenamefont
  {{Aloisio}}}]{2020arXiv200701302E}%
  \BibitemOpen
  \bibfield  {author} {\bibinfo {author} {\bibfnamefont {C.}~\bibnamefont
  {{Evoli}}}, \bibinfo {author} {\bibfnamefont {P.}~\bibnamefont {{Blasi}}},
  \bibinfo {author} {\bibfnamefont {E.}~\bibnamefont {{Amato}}},\ and\ \bibinfo
  {author} {\bibfnamefont {R.}~\bibnamefont {{Aloisio}}},\ }\bibfield  {title}
  {\bibinfo {title} {{The signature of energy losses on the cosmic ray electron
  spectrum}},\ }\href@noop {} {\bibfield  {journal} {\bibinfo  {journal} {arXiv
  e-prints}\ ,\ \bibinfo {eid} {arXiv:2007.01302}} (\bibinfo {year}
  {2020}{\natexlab{a}})},\ \Eprint {https://arxiv.org/abs/2007.01302}
  {arXiv:2007.01302 [astro-ph.HE]} \BibitemShut {NoStop}%
\bibitem [{\citenamefont {Aguilar}\ \emph
  {et~al.}(2014{\natexlab{b}})\citenamefont {Aguilar} \emph
  {et~al.}}]{Aguilar:2014fea}%
  \BibitemOpen
  \bibfield  {author} {\bibinfo {author} {\bibfnamefont {M.}~\bibnamefont
  {Aguilar}} \emph {et~al.} (\bibinfo {collaboration} {AMS}),\ }\bibfield
  {title} {\bibinfo {title} {{Precision Measurement of the all-lepton Flux in
  Primary Cosmic Rays from 0.5 GeV to 1 TeV with the Alpha Magnetic
  Spectrometer on the International Space Station}},\ }\href
  {https://doi.org/10.1103/PhysRevLett.113.221102} {\bibfield  {journal}
  {\bibinfo  {journal} {Phys. Rev. Lett.}\ }\textbf {\bibinfo {volume} {113}},\
  \bibinfo {pages} {221102} (\bibinfo {year} {2014}{\natexlab{b}})}\BibitemShut
  {NoStop}%
\bibitem [{\citenamefont {Vink}(2012)}]{Vink:2011ei}%
  \BibitemOpen
  \bibfield  {author} {\bibinfo {author} {\bibfnamefont {J.}~\bibnamefont
  {Vink}},\ }\bibfield  {title} {\bibinfo {title} {{Supernova remnants: the
  X-ray perspective}},\ }\href {https://doi.org/10.1007/s00159-011-0049-1}
  {\bibfield  {journal} {\bibinfo  {journal} {Astron. Astrophys. Rev.}\
  }\textbf {\bibinfo {volume} {20}},\ \bibinfo {pages} {1} (\bibinfo {year}
  {2012})},\ \Eprint {https://arxiv.org/abs/1112.0576} {arXiv:1112.0576
  [astro-ph.HE]} \BibitemShut {NoStop}%
\bibitem [{\citenamefont {Diesing}\ and\ \citenamefont
  {Caprioli}(2019)}]{Diesing:2019lwm}%
  \BibitemOpen
  \bibfield  {author} {\bibinfo {author} {\bibfnamefont {R.}~\bibnamefont
  {Diesing}}\ and\ \bibinfo {author} {\bibfnamefont {D.}~\bibnamefont
  {Caprioli}},\ }\bibfield  {title} {\bibinfo {title} {{Spectrum of Electrons
  Accelerated in Supernova Remnants}},\ }\href
  {https://doi.org/10.1103/PhysRevLett.123.071101} {\bibfield  {journal}
  {\bibinfo  {journal} {Phys. Rev. Lett.}\ }\textbf {\bibinfo {volume} {123}},\
  \bibinfo {pages} {071101} (\bibinfo {year} {2019})},\ \Eprint
  {https://arxiv.org/abs/1905.07414} {arXiv:1905.07414 [astro-ph.HE]}
  \BibitemShut {NoStop}%
%%CITATION = ARXIV:1905.07414;%%
\bibitem [{\citenamefont {Aguilar}\ \emph
  {et~al.}(2015{\natexlab{b}})\citenamefont {Aguilar} \emph
  {et~al.}}]{Aguilar:2015ooa}%
  \BibitemOpen
  \bibfield  {author} {\bibinfo {author} {\bibfnamefont {M.}~\bibnamefont
  {Aguilar}} \emph {et~al.} (\bibinfo {collaboration} {AMS}),\ }\bibfield
  {title} {\bibinfo {title} {{Precision Measurement of the Proton Flux in
  Primary Cosmic Rays from Rigidity 1 GV to 1.8 TV with the Alpha Magnetic
  Spectrometer on the International Space Station}},\ }\href
  {https://doi.org/10.1103/PhysRevLett.114.171103} {\bibfield  {journal}
  {\bibinfo  {journal} {Phys. Rev. Lett.}\ }\textbf {\bibinfo {volume} {114}},\
  \bibinfo {pages} {171103} (\bibinfo {year} {2015}{\natexlab{b}})}\BibitemShut
  {NoStop}%
\bibitem [{\citenamefont {Tatischeff}(2009)}]{Tatischeff:2009kh}%
  \BibitemOpen
  \bibfield  {author} {\bibinfo {author} {\bibfnamefont {V.}~\bibnamefont
  {Tatischeff}},\ }\bibfield  {title} {\bibinfo {title} {{Radio emission and
  nonlinear diffusive shock acceleration of cosmic rays in the supernova SN
  1993J}},\ }\href {https://doi.org/10.1051/0004-6361/200811511} {\bibfield
  {journal} {\bibinfo  {journal} {Astron. Astrophys.}\ }\textbf {\bibinfo
  {volume} {499}},\ \bibinfo {pages} {191} (\bibinfo {year} {2009})},\ \Eprint
  {https://arxiv.org/abs/0903.2944} {arXiv:0903.2944 [astro-ph.HE]}
  \BibitemShut {NoStop}%
\bibitem [{\citenamefont {Bell}(2013)}]{Bell:2013vxa}%
  \BibitemOpen
  \bibfield  {author} {\bibinfo {author} {\bibfnamefont {A.}~\bibnamefont
  {Bell}},\ }\bibfield  {title} {\bibinfo {title} {{Cosmic ray acceleration}},\
  }\href {https://doi.org/10.1016/j.astropartphys.2012.05.022} {\bibfield
  {journal} {\bibinfo  {journal} {Astropart. Phys.}\ }\textbf {\bibinfo
  {volume} {43}},\ \bibinfo {pages} {56} (\bibinfo {year} {2013})}\BibitemShut
  {NoStop}%
\bibitem [{\citenamefont {Zirakashvili}\ and\ \citenamefont
  {Ptuskin}(2017)}]{Zirakashvili2017}%
  \BibitemOpen
  \bibfield  {author} {\bibinfo {author} {\bibfnamefont {V.~N.}\ \bibnamefont
  {Zirakashvili}}\ and\ \bibinfo {author} {\bibfnamefont {V.~S.}\ \bibnamefont
  {Ptuskin}},\ }\bibfield  {title} {\bibinfo {title} {Acceleration of particles
  and generation of nonthermal emission in old supernova remnants},\ }\href
  {https://doi.org/10.3103/S1062873817040426} {\bibfield  {journal} {\bibinfo
  {journal} {Bulletin of the Russian Academy of Sciences: Physics}\ }\textbf
  {\bibinfo {volume} {81}},\ \bibinfo {pages} {434} (\bibinfo {year}
  {2017})}\BibitemShut {NoStop}%
\bibitem [{\citenamefont {Ahlers}\ and\ \citenamefont
  {Mertsch}(2017)}]{Ahlers:2016rox}%
  \BibitemOpen
  \bibfield  {author} {\bibinfo {author} {\bibfnamefont {M.}~\bibnamefont
  {Ahlers}}\ and\ \bibinfo {author} {\bibfnamefont {P.}~\bibnamefont
  {Mertsch}},\ }\bibfield  {title} {\bibinfo {title} {{Origin of Small-Scale
  Anisotropies in Galactic Cosmic Rays}},\ }\href
  {https://doi.org/10.1016/j.ppnp.2017.01.004} {\bibfield  {journal} {\bibinfo
  {journal} {Prog. Part. Nucl. Phys.}\ }\textbf {\bibinfo {volume} {94}},\
  \bibinfo {pages} {184} (\bibinfo {year} {2017})},\ \Eprint
  {https://arxiv.org/abs/1612.01873} {arXiv:1612.01873 [astro-ph.HE]}
  \BibitemShut {NoStop}%
\bibitem [{\citenamefont {{Ginzburg}}\ and\ \citenamefont
  {{Syrovatskii}}(1964)}]{Ginzburg1964}%
  \BibitemOpen
  \bibfield  {author} {\bibinfo {author} {\bibfnamefont {V.~L.}\ \bibnamefont
  {{Ginzburg}}}\ and\ \bibinfo {author} {\bibfnamefont {S.~I.}\ \bibnamefont
  {{Syrovatskii}}},\ }\href@noop {} {\emph {\bibinfo {title} {The Origin of
  Cosmic Rays, New York: Macmillan, 1964}}}\ (\bibinfo  {publisher} {Pergamon
  Press},\ \bibinfo {year} {1964})\BibitemShut {NoStop}%
\bibitem [{\citenamefont {Hunter}\ \emph {et~al.}(1997)\citenamefont {Hunter}
  \emph {et~al.}}]{Hunger:1997we}%
  \BibitemOpen
  \bibfield  {author} {\bibinfo {author} {\bibfnamefont {S.}~\bibnamefont
  {Hunter}} \emph {et~al.},\ }\bibfield  {title} {\bibinfo {title} {{EGRET
  observations of the diffuse gamma-ray emission from the galactic plane}},\
  }\href {https://doi.org/10.1086/304012} {\bibfield  {journal} {\bibinfo
  {journal} {Astrophys. J.}\ }\textbf {\bibinfo {volume} {481}},\ \bibinfo
  {pages} {205} (\bibinfo {year} {1997})}\BibitemShut {NoStop}%
\bibitem [{\citenamefont {Bartoli}\ \emph {et~al.}(2015)\citenamefont {Bartoli}
  \emph {et~al.}}]{Bartoli:2015ysa}%
  \BibitemOpen
  \bibfield  {author} {\bibinfo {author} {\bibfnamefont {B.}~\bibnamefont
  {Bartoli}} \emph {et~al.} (\bibinfo {collaboration} {ARGO-YBJ}),\ }\bibfield
  {title} {\bibinfo {title} {{Argo-ybj Observation of the Large-scale Cosmic
  ray Anisotropy During the Solar Minimum Between Cycles 23 and 24}},\ }\href
  {https://doi.org/10.1088/0004-637X/809/1/90} {\bibfield  {journal} {\bibinfo
  {journal} {Astrophys. J.}\ }\textbf {\bibinfo {volume} {809}},\ \bibinfo
  {pages} {90} (\bibinfo {year} {2015})}\BibitemShut {NoStop}%
\bibitem [{\citenamefont {Bartoli}\ \emph {et~al.}(2018)\citenamefont {Bartoli}
  \emph {et~al.}}]{Bartoli:2018ach}%
  \BibitemOpen
  \bibfield  {author} {\bibinfo {author} {\bibfnamefont {B.}~\bibnamefont
  {Bartoli}} \emph {et~al.} (\bibinfo {collaboration} {ARGO-YBJ}),\ }\bibfield
  {title} {\bibinfo {title} {{Galactic Cosmic-Ray Anisotropy in the Northern
  Hemisphere from the ARGO-YBJ Experiment during 2008--2012}},\ }\href
  {https://doi.org/10.3847/1538-4357/aac6cc} {\bibfield  {journal} {\bibinfo
  {journal} {Astrophys. J.}\ }\textbf {\bibinfo {volume} {861}},\ \bibinfo
  {pages} {93} (\bibinfo {year} {2018})},\ \Eprint
  {https://arxiv.org/abs/1805.08980} {arXiv:1805.08980 [astro-ph.HE]}
  \BibitemShut {NoStop}%
\bibitem [{\citenamefont {Amenomori}(2017)}]{Amenomori:2017jbv}%
  \BibitemOpen
  \bibfield  {author} {\bibinfo {author} {\bibfnamefont {M.}~\bibnamefont
  {Amenomori}} (\bibinfo {collaboration} {Tibet AS-gamma}),\ }\bibfield
  {title} {\bibinfo {title} {{Northern sky Galactic Cosmic Ray anisotropy
  between 10-1000 TeV with the Tibet Air Shower Array}},\ }\href
  {https://doi.org/10.3847/1538-4357/836/2/153} {\bibfield  {journal} {\bibinfo
   {journal} {Astrophys. J.}\ }\textbf {\bibinfo {volume} {836}},\ \bibinfo
  {pages} {153} (\bibinfo {year} {2017})},\ \Eprint
  {https://arxiv.org/abs/1701.07144} {arXiv:1701.07144 [astro-ph.HE]}
  \BibitemShut {NoStop}%
\bibitem [{\citenamefont {{Evoli}}\ \emph
  {et~al.}(2020{\natexlab{b}})\citenamefont {{Evoli}}, \citenamefont {{Amato}},
  \citenamefont {{Blasi}},\ and\ \citenamefont
  {{Aloisio}}}]{2020arXiv201011955E}%
  \BibitemOpen
  \bibfield  {author} {\bibinfo {author} {\bibfnamefont {C.}~\bibnamefont
  {{Evoli}}}, \bibinfo {author} {\bibfnamefont {E.}~\bibnamefont {{Amato}}},
  \bibinfo {author} {\bibfnamefont {P.}~\bibnamefont {{Blasi}}},\ and\ \bibinfo
  {author} {\bibfnamefont {R.}~\bibnamefont {{Aloisio}}},\ }\bibfield  {title}
  {\bibinfo {title} {{Galactic factories of cosmic-ray electrons and
  positrons}},\ }\href@noop {} {\bibfield  {journal} {\bibinfo  {journal}
  {arXiv e-prints}\ ,\ \bibinfo {eid} {arXiv:2010.11955}} (\bibinfo {year}
  {2020}{\natexlab{b}})},\ \Eprint {https://arxiv.org/abs/2010.11955}
  {arXiv:2010.11955 [astro-ph.HE]} \BibitemShut {NoStop}%
\bibitem [{\citenamefont {{Berezhko}}\ \emph {et~al.}(1994)\citenamefont
  {{Berezhko}}, \citenamefont {{Yelshin}},\ and\ \citenamefont
  {{Ksenofontov}}}]{berezhko94}%
  \BibitemOpen
  \bibfield  {author} {\bibinfo {author} {\bibfnamefont {E.~G.}\ \bibnamefont
  {{Berezhko}}}, \bibinfo {author} {\bibfnamefont {V.~K.}\ \bibnamefont
  {{Yelshin}}},\ and\ \bibinfo {author} {\bibfnamefont {L.~T.}\ \bibnamefont
  {{Ksenofontov}}},\ }\bibfield  {title} {\bibinfo {title} {{Numerical
  investigation of cosmic ray acceleration in supernova remnants}},\ }\href
  {https://doi.org/10.1016/0927-6505(94)90043-4} {\bibfield  {journal}
  {\bibinfo  {journal} {Astroparticle Physics}\ }\textbf {\bibinfo {volume}
  {2}},\ \bibinfo {pages} {215} (\bibinfo {year} {1994})}\BibitemShut {NoStop}%
\bibitem [{\citenamefont {{Schure}}\ and\ \citenamefont
  {{Bell}}(2013)}]{schure13}%
  \BibitemOpen
  \bibfield  {author} {\bibinfo {author} {\bibfnamefont {K.~M.}\ \bibnamefont
  {{Schure}}}\ and\ \bibinfo {author} {\bibfnamefont {A.~R.}\ \bibnamefont
  {{Bell}}},\ }\bibfield  {title} {\bibinfo {title} {{Cosmic ray acceleration
  in young supernova remnants}},\ }\href
  {https://doi.org/10.1093/mnras/stt1371} {\bibfield  {journal} {\bibinfo
  {journal} {\mnras}\ }\textbf {\bibinfo {volume} {435}},\ \bibinfo {pages}
  {1174} (\bibinfo {year} {2013})},\ \Eprint {https://arxiv.org/abs/1307.6575}
  {arXiv:1307.6575 [astro-ph.HE]} \BibitemShut {NoStop}%
\bibitem [{\citenamefont {{Bell}}(2004)}]{bell04}%
  \BibitemOpen
  \bibfield  {author} {\bibinfo {author} {\bibfnamefont {A.~R.}\ \bibnamefont
  {{Bell}}},\ }\bibfield  {title} {\bibinfo {title} {{Turbulent amplification
  of magnetic field and diffusive shock acceleration of cosmic rays}},\ }\href
  {https://doi.org/10.1111/j.1365-2966.2004.08097.x} {\bibfield  {journal}
  {\bibinfo  {journal} {\mnras}\ }\textbf {\bibinfo {volume} {353}},\ \bibinfo
  {pages} {550} (\bibinfo {year} {2004})}\BibitemShut {NoStop}%
\bibitem [{\citenamefont {{Truelove}}\ and\ \citenamefont
  {{McKee}}(1999)}]{truelove99}%
  \BibitemOpen
  \bibfield  {author} {\bibinfo {author} {\bibfnamefont {J.~K.}\ \bibnamefont
  {{Truelove}}}\ and\ \bibinfo {author} {\bibfnamefont {C.~F.}\ \bibnamefont
  {{McKee}}},\ }\bibfield  {title} {\bibinfo {title} {{Evolution of
  Nonradiative Supernova Remnants}},\ }\href {https://doi.org/10.1086/313176}
  {\bibfield  {journal} {\bibinfo  {journal} {\apjs}\ }\textbf {\bibinfo
  {volume} {120}},\ \bibinfo {pages} {299} (\bibinfo {year}
  {1999})}\BibitemShut {NoStop}%
\bibitem [{\citenamefont {{Cioffi}}\ \emph {et~al.}(1988)\citenamefont
  {{Cioffi}}, \citenamefont {{McKee}},\ and\ \citenamefont
  {{Bertschinger}}}]{cioffi88}%
  \BibitemOpen
  \bibfield  {author} {\bibinfo {author} {\bibfnamefont {D.~F.}\ \bibnamefont
  {{Cioffi}}}, \bibinfo {author} {\bibfnamefont {C.~F.}\ \bibnamefont
  {{McKee}}},\ and\ \bibinfo {author} {\bibfnamefont {E.}~\bibnamefont
  {{Bertschinger}}},\ }\bibfield  {title} {\bibinfo {title} {{Dynamics of
  Radiative Supernova Remnants}},\ }\href {https://doi.org/10.1086/166834}
  {\bibfield  {journal} {\bibinfo  {journal} {\apj}\ }\textbf {\bibinfo
  {volume} {334}},\ \bibinfo {pages} {252} (\bibinfo {year}
  {1988})}\BibitemShut {NoStop}%
\bibitem [{\citenamefont {{Berezhko}}(1996)}]{1996APh.....5..367B}%
  \BibitemOpen
  \bibfield  {author} {\bibinfo {author} {\bibfnamefont {E.~G.}\ \bibnamefont
  {{Berezhko}}},\ }\bibfield  {title} {\bibinfo {title} {{Maximum energy of
  cosmic rays accelerated by supernova shocks}},\ }\href
  {https://doi.org/10.1016/0927-6505(96)00037-0} {\bibfield  {journal}
  {\bibinfo  {journal} {Astroparticle Physics}\ }\textbf {\bibinfo {volume}
  {5}},\ \bibinfo {pages} {367} (\bibinfo {year} {1996})}\BibitemShut {NoStop}%
\bibitem [{\citenamefont {{McKee}}(1982)}]{1982ASIC...90..433M}%
  \BibitemOpen
  \bibfield  {author} {\bibinfo {author} {\bibfnamefont {C.~F.}\ \bibnamefont
  {{McKee}}},\ }\bibfield  {title} {\bibinfo {title} {{The evolution of
  supernova remnants and the structure of the interstellar medium}},\ }in\
  \href@noop {} {\emph {\bibinfo {booktitle} {Supernovae: A Survey of Current
  Research}}},\ \bibinfo {series} {NATO Advanced Study Institute (ASI) Series
  C}, Vol.~\bibinfo {volume} {90},\ \bibinfo {editor} {edited by\ \bibinfo
  {editor} {\bibfnamefont {M.~J.}\ \bibnamefont {{Rees}}}\ and\ \bibinfo
  {editor} {\bibfnamefont {R.~J.}\ \bibnamefont {{Stoneham}}}}\ (\bibinfo
  {year} {1982})\ pp.\ \bibinfo {pages} {433--457}\BibitemShut {NoStop}%
\bibitem [{\citenamefont {{V{\"o}lk}}\ \emph {et~al.}(2005)\citenamefont
  {{V{\"o}lk}}, \citenamefont {{Berezhko}},\ and\ \citenamefont
  {{Ksenofontov}}}]{voelk05}%
  \BibitemOpen
  \bibfield  {author} {\bibinfo {author} {\bibfnamefont {H.~J.}\ \bibnamefont
  {{V{\"o}lk}}}, \bibinfo {author} {\bibfnamefont {E.~G.}\ \bibnamefont
  {{Berezhko}}},\ and\ \bibinfo {author} {\bibfnamefont {L.~T.}\ \bibnamefont
  {{Ksenofontov}}},\ }\bibfield  {title} {\bibinfo {title} {{Magnetic field
  amplification in Tycho and other shell-type supernova remnants}},\ }\href
  {https://doi.org/10.1051/0004-6361:20042015} {\bibfield  {journal} {\bibinfo
  {journal} {\aap}\ }\textbf {\bibinfo {volume} {433}},\ \bibinfo {pages} {229}
  (\bibinfo {year} {2005})},\ \Eprint {https://arxiv.org/abs/astro-ph/0409453}
  {arXiv:astro-ph/0409453 [astro-ph]} \BibitemShut {NoStop}%
\bibitem [{\citenamefont {{Ohira}}\ \emph {et~al.}(2012)\citenamefont
  {{Ohira}}, \citenamefont {{Yamazaki}}, \citenamefont {{Kawanaka}},\ and\
  \citenamefont {{Ioka}}}]{ohira12}%
  \BibitemOpen
  \bibfield  {author} {\bibinfo {author} {\bibfnamefont {Y.}~\bibnamefont
  {{Ohira}}}, \bibinfo {author} {\bibfnamefont {R.}~\bibnamefont {{Yamazaki}}},
  \bibinfo {author} {\bibfnamefont {N.}~\bibnamefont {{Kawanaka}}},\ and\
  \bibinfo {author} {\bibfnamefont {K.}~\bibnamefont {{Ioka}}},\ }\bibfield
  {title} {\bibinfo {title} {{Escape of cosmic-ray electrons from supernova
  remnants}},\ }\href {https://doi.org/10.1111/j.1365-2966.2012.21908.x}
  {\bibfield  {journal} {\bibinfo  {journal} {\mnras}\ }\textbf {\bibinfo
  {volume} {427}},\ \bibinfo {pages} {91} (\bibinfo {year} {2012})},\ \Eprint
  {https://arxiv.org/abs/1106.1810} {arXiv:1106.1810 [astro-ph.HE]}
  \BibitemShut {NoStop}%
\bibitem [{\citenamefont {{Parizot}}\ \emph {et~al.}(2006)\citenamefont
  {{Parizot}}, \citenamefont {{Marcowith}}, \citenamefont {{Ballet}},\ and\
  \citenamefont {{Gallant}}}]{parizot06}%
  \BibitemOpen
  \bibfield  {author} {\bibinfo {author} {\bibfnamefont {E.}~\bibnamefont
  {{Parizot}}}, \bibinfo {author} {\bibfnamefont {A.}~\bibnamefont
  {{Marcowith}}}, \bibinfo {author} {\bibfnamefont {J.}~\bibnamefont
  {{Ballet}}},\ and\ \bibinfo {author} {\bibfnamefont {Y.~A.}\ \bibnamefont
  {{Gallant}}},\ }\bibfield  {title} {\bibinfo {title} {{Observational
  constraints on energetic particle diffusion in young supernovae remnants:
  amplified magnetic field and maximum energy}},\ }\href
  {https://doi.org/10.1051/0004-6361:20064985} {\bibfield  {journal} {\bibinfo
  {journal} {\aap}\ }\textbf {\bibinfo {volume} {453}},\ \bibinfo {pages} {387}
  (\bibinfo {year} {2006})},\ \Eprint {https://arxiv.org/abs/astro-ph/0603723}
  {arXiv:astro-ph/0603723 [astro-ph]} \BibitemShut {NoStop}%
\bibitem [{\citenamefont {{Pelgrims}}\ \emph {et~al.}(2020)\citenamefont
  {{Pelgrims}}, \citenamefont {{Ferri{\`e}re}}, \citenamefont {{Boulanger}},
  \citenamefont {{Lallement}},\ and\ \citenamefont
  {{Montier}}}]{2020A&A...636A..17P}%
  \BibitemOpen
  \bibfield  {author} {\bibinfo {author} {\bibfnamefont {V.}~\bibnamefont
  {{Pelgrims}}}, \bibinfo {author} {\bibfnamefont {K.}~\bibnamefont
  {{Ferri{\`e}re}}}, \bibinfo {author} {\bibfnamefont {F.}~\bibnamefont
  {{Boulanger}}}, \bibinfo {author} {\bibfnamefont {R.}~\bibnamefont
  {{Lallement}}},\ and\ \bibinfo {author} {\bibfnamefont {L.}~\bibnamefont
  {{Montier}}},\ }\bibfield  {title} {\bibinfo {title} {{Modeling the
  magnetized Local Bubble from dust data}},\ }\href
  {https://doi.org/10.1051/0004-6361/201937157} {\bibfield  {journal} {\bibinfo
   {journal} {\aap}\ }\textbf {\bibinfo {volume} {636}},\ \bibinfo {eid} {A17}
  (\bibinfo {year} {2020})},\ \Eprint {https://arxiv.org/abs/1911.09691}
  {arXiv:1911.09691 [astro-ph.GA]} \BibitemShut {NoStop}%
\end{thebibliography}%
\end{document}